\documentclass[prd,nofootinbib,onecolumn,preprint,10pt]{revtex4}

\usepackage{graphicx}
\usepackage{bm}
\usepackage{float}
\usepackage{subfigure}
\usepackage{amsmath}
\usepackage{amsfonts}
\usepackage{color}
\usepackage{slashed}
\usepackage{fancyhdr}
\usepackage[colorlinks=true,linkcolor=black,citecolor=black,filecolor=black,urlcolor=black,unicode]{hyperref}

\newcommand{\GN}{G_{\rm N}}

\newcommand{\dd}{\text{d}}

\begin{document}
	
\title{Electromagnetic and Gravitational Radiation in All Dimensions: \\ A Classical Field Theory Treatment}
	
\author{Yi-Zen Chu$^{1,2}$}
\affiliation{
$\,^1$Department of Physics, National Central University, Chungli 32001, Taiwan \\
$\,^2$Center for High Energy and High Field Physics (CHiP), National Central University, Chungli 32001, Taiwan
}

\begin{abstract}
\noindent How long does a light bulb shine in odd dimensional flat spacetimes, according to a distant observer? This question is non-trivial because electromagnetic and gravitational waves, despite being comprised of massless particles, can develop tails: they travel inside the light cone. To this end, I attempt to close a gap in the literature by first deriving, strictly within classical field theory, the real-time electromagnetic dipole and gravitational quadrupole energy and angular momentum radiation formulas in all relevant dimensions. The even dimensional case, where massless signals travel strictly on the null cone, depends on the time derivatives of the dipoles and quadrupoles solely at retarded time; whereas the odd dimensional ones involve an integral over their retarded histories. Despite the propagation of light inside the null cone, however, I argue that a monochromatic light bulb of some intrinsic duration in odd dimensions remains approximately the same apparent duration to a distant detector, though the tail effect does produce a phase shift and adds to the signal several transitory non-oscillatory inverse square roots in time. Analogous remarks apply to a distant gravitational wave detector hearing from a finite duration quasi-periodic quadrupole source. 
\end{abstract}

\maketitle


\section{Motivation}
\label{Section_Introduction}

\begin{figure}[!ht]
	\begin{center}
		\includegraphics[width=5in]{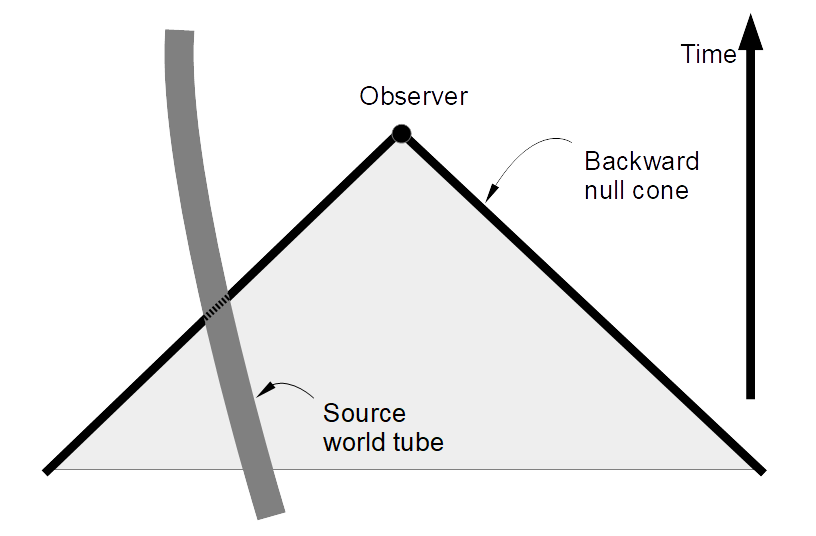}
		\caption{Causal structure of signals in odd versus even dimensional Minkowski spacetime. The Dirac delta function of eq. \eqref{G_EvenDim}, when employed in eq. \eqref{MasslessScalar_Solution}, tells us the signal received by the observer can only come from her past light cone. This is the dashed segment, representing the intersection between the observer's backward light cone and the world tube of the source responsible for the waves. In contrast, the step function of eq. \eqref{G_OddDim} when employed in eq. \eqref{MasslessScalar_Solution} says the signal detected by the observer comes from {\it within} her null cone (gray region). This means the signal she receives at a given moment is the superposition of the waves emitted from the source's world tube lying within this gray region.}
		\label{BackwardNullCone}
	\end{center}
\end{figure}

The flat spacetime wave operator $\partial^2$ in even dimensions higher than two ($d \geq 4$) admits the retarded Green's function
\begin{align}
\label{G_EvenDim}
G_{\text{even $d \geq 4$}}[x-x']
&= \left(-\frac{1}{2\pi R}\frac{\partial}{\partial R}\right)^{\frac{d-4}{2}} \frac{\delta[T-R]}{4\pi R} , \\
T&\equiv t-t', \qquad R \equiv |\vec{x}-\vec{x}'| ;
\end{align}
while in all odd dimensions ($d \geq 3$), it admits instead the retarded Green's function
\begin{align}
\label{G_OddDim}
G_{\text{odd $d \geq 3$}}[x-x']
&= \left(-\frac{1}{2\pi R}\frac{\partial}{\partial R}\right)^{\frac{d-3}{2}} \frac{\Theta[T-R]}{2\pi \sqrt{T^2-R^2}} .
\end{align}
With the Minkowski metric given by $\eta_{\mu\nu} \equiv \text{diag}[1,-1,\dots,-1]$, both Green's functions in equations \eqref{G_EvenDim} and \eqref{G_OddDim} obey the wave equation
\begin{align}
\label{G_WaveEquation}
\partial_x^2 G_d[x-x'] = \partial_{x'}^2 G_d[x-x'] = \delta^{(d)}[x-x'] ,
\end{align}
where $\partial^2 \equiv \eta^{\alpha\beta} \partial_\alpha \partial_\beta$ and $\delta^{(d)}$ is the $d-$dimensional Dirac delta function. By viewing $x^\mu = (t,\vec{x})$ as the location of some observer, and $x'^\mu = (t',\vec{x}')$ as that of the source, eq. \eqref{G_WaveEquation} tells us the retarded Green's function is the signal at $x$ generated by a spacetime point source at $x'$. Moreover, the Dirac $\delta-$function in eq. \eqref{G_EvenDim} tells us (massless) waves propagate strictly on the null cone $T=R$ in even dimensions $d \geq 4$; whereas the Heaviside step function in eq. \eqref{G_OddDim} informs us the odd dimensional counterparts develop tails -- namely, the signal permeates the interior of the light cone $T > R$. To further elaborate this distinction between the causal structure of the wave signals in odd versus even dimensions, consider the massless scalar wave equation
\begin{align}
	\label{MasslessScalar_WaveEqn}
	\partial^2 \psi = J ,
\end{align}
whose solution can be expressed as the convolution
\begin{align}
	\label{MasslessScalar_Solution}
	\psi[x] = \int_{\mathbb{R}^{1,d-1}} G_d[x-x'] J[x'] \dd^d x' .
\end{align}
As I illustrate in Fig. \eqref{BackwardNullCone}, due to the Dirac delta function of eq. \eqref{G_EvenDim} enforcing null cone only signal transmission, the observer receives waves emitted only from the intersection between her past null cone with the source's world tube (dashed line). Whereas, in odd dimensions, due to the tail transmission of signals arising from the step function of eq. \eqref{G_OddDim}, the observer receives waves emitted from the entire past history of the source's world tube that lies within the interior of her past null cone.

One of the primary motivations behind the current work is, the {\it real-time} gravitational quadrupole formula -- the leading non-relativistic expression for power emitted by a gravitational system -- is not known in odd dimensions $d \geq 5$.\footnote{It should be noted, however, that the real-time massless scalar synchrotron radiation in odd 2+1 and 4+1 dimensions has been computed by Gal'tsov and Khlopunov \cite{Galtsov:2020hhn}.} Specifically, Cardoso, Dias and Lemos \cite{Cardoso:2002pa} and Cardoso, Dias and Figueras \cite{Cardoso:2008gn} encountered difficulties deriving it because of the tail effect. In the latter \cite{Cardoso:2008gn}, the gravitational quadrupole formula was in fact obtained in all dimensions, but only in frequency space; moreover, it was derived by employing quantum field theory techniques to extract the answer from the vacuum-to-vacuum transition amplitude. That such sophisticated quantum methods have to be brought in to derive an incomplete classical result, where the real-time information is lost, suggests more effort is warranted to improve our physical understanding of this basic result in gravitation.

But why bother to understand better the quadrupole formula in a dimension other than the $3+1$ ones we reside in? One reason is that $d$, the spacetime dimension, may be considered a parameter in the equations of physics; and by varying it we may develop a deeper appreciation of the phenomenon that follow from them peculiar to our 4D world. For instance, although null-traveling plane wave solutions of the homogeneous equation $\partial^2 \psi = 0$ exist in all dimensions, namely $\psi(x) = \exp(i \vec{k} \cdot \vec{x} \pm i |\vec{k}|t)$, the signals produced by a physically isolated system arises from the superposition of the Green's function weighted by the associated source(s) and is therefore only tail-free in even dimensions $d \geq 4$. Since Green's functions are $s-$waves with respect to the variables $T$ and $R$, there therefore appears to be a fundamental incompatibility between spherical waves and strictly-null propagation of physical signals in odd dimensions.

Four dimensions is also the only dimension where both radiation and the static Coulomb and Newtonian potentials appear at the same order in the asymptotic $1/r$ expansion, where $r$ is the radial distance between observer and source. To see this, first recall that, in an inertial Lorentz frame, the spatial volume measure in spherical coordinates is $r^{d-2} \dd r \dd\Omega$, where $\dd\Omega$ is the infinitesimal solid angle. If $T^{\mu\nu}$ is the (pseudo-)energy-momentum-shear-stress tensor of the associated radiation and $\widehat{r}$ is the unit radial vector, the energy flux is
\begin{align}
	\label{EnergyFlux}
	\frac{\dd E}{\dd t\Omega} = \lim_{r\to\infty} r^{d-2} T^{0i} \widehat{r}^i .
\end{align}
\footnote{Throughout this paper, Latin/English alphabets run over spatial indices whereas Greek ones run over spacetime, with the $0$th component denoting the time coordinate. Einstein summation is in force unless otherwise stated.}Hence, to yield a well defined non-zero energy loss to infinity, the radiative portions of $T^{0i}$ must scale as $1/r^{d-2}$. As we shall witness below, because $T^{\mu\nu}$ for both electromagnetism and gravitation are quadratic in the fields, the radiative terms of the photon and graviton field must in turn scale as $1/r^{(d/2)-1}$. On the other hand, in the static limit, both photon and graviton fields must satisfy some version of Gauss' law. For a point charge or mass, the corresponding linearized field is the Green's function of the Laplacian, which scales as $1/r^{d-3}$. (The special case of $d=3$ does not even return a power law Green's function, but rather a logarithm: $1/\vec{\nabla}^2 = (2\pi)^{-1}\ln r$.) The unique solution to $1/r^{(d/2)-1} = 1/r^{d-3}$ is $d=4$. By taking the far zone limits of the solutions to the Lorenz gauge vector potential and the linear de Donder gauge gravitational metric perturbations in, respectively, equations (A74)-(A75) and (B11)-(B12) of \cite{Chu:2016ngc}, we further see that the coefficients of $1/r^{(d/2)-1}$ are proportional to (in odd dimensions) $(d-3)/2$ and (in even dimensions) $(d-4)/2$ time derivatives of the associated electromagnetic current and matter stress tensor; whereas that of the $1/r^{d-3}$ has zero time derivatives acting on these same sources. To this end, in \S \eqref{Section_EM} and \S \eqref{Section_Gravity} below, we will not only compute the electromagnetic and gravitational energy flux given in eq. \eqref{EnergyFlux}, but also their angular momentum radiated. By considering rotations on the spatial $(i,j)-$plane, the associated angular momentum flux is
\begin{align}
	\label{AngularMomentumFlux}
	\frac{\dd L^{ij}}{\dd t \dd \Omega} = \lim_{r\to\infty} r^{d-2} x^{[i} T^{j]k} \widehat{r}^k .
\end{align}
Studying physics in Minkowski spacetime of various dimensions may also provide us with a simpler context to study not only {\it why} the tail effect exists but also {\it how} to analyze and compute its consequences. This could potentially lead to novel methods or insights into the curved spacetime scenario, where understanding the tail induced self-force acting upon compact objects orbiting supermassive black holes at the center of galaxies is important (at least within the de Donder gauge) for modeling gravitational waves from such Extreme-Mass-Ratio-Inspiral (EMRI) systems -- a prime target for the upcoming space-based LISA mission.

The contents of this paper are as follows. As a warm-up to the technically more arduous gravitation calculation, in \S \eqref{Section_EM}, I shall first generalize the leading order non-relativistic electromagnetic dipole radiation formula in 4 dimensions, namely the Heaviside-Lorentz and $c = 1$ natural units\footnote{Which I shall deploy throughout the rest of this paper.} based expression
\begin{align}
\left. \frac{\dd E}{\dd t} \right\vert_{d=4} = \frac{(\partial_t^2 \vec{d}[t-r])^2}{6\pi} ,
\end{align}
to all dimensions ($d \geq 3$) greater than two. Throughout, I shall denote the dipole moment as $\vec{d}$; and will always assume the spatial origin $\vec{0}$ to be located somewhere within the electromagnetic current or matter energy-momentum tensor, so that $t-r$ is the retarded time. In even dimensions $d \geq 4$, we will discover below
\begin{align}
\label{Maxwell_DipoleEnergyRadiation_EvenDim}
\frac{\dd E}{\dd t}
&= \frac{d-2}{2^{d} \pi^{\frac{d-3}{2}} \Gamma[\frac{d+1}{2}]} \left( \partial_t^{\frac{d}{2}} \vec{d}[t-r] \right)^2 ,
\end{align}
where $\Gamma$ is the Gamma function and the square of a multi-component object will always mean the Euclidean dot product with itself (e.g., $\vec{a}^2 \equiv \vec{a}\cdot\vec{a}$). On the other hand, in odd dimensions $d \geq 3$,
\begin{align}
\label{Maxwell_DipoleEnergyRadiation_OddDim}
\frac{\dd E}{\dd t}
	&= \frac{d-2}{2^d \pi^{\frac{d-1}{2}} \Gamma[\frac{d+1}{2}]} \left( \int_{0}^{\infty} \frac{\dd\mu}{\mu^{\frac{1}{2}}} \partial_t^{\frac{d+1}{2}} \vec{d}[t-r-\mu] \right)^2 .
\end{align}
The retarded history integral over $\mu \in [0,\infty)$ sums the $(1/2)(d+1)$th time derivative of the dipole's contribution to the electromagnetic power from retarded time $t-r$ to past infinity $\lim_{\mu \to \infty}(t-r-\mu)$. Its presence is the direct result of the tail effect, for since massless waves do travel inside the null cone in odd dimensions, the observer is now sensitive to the entire past history of the relevant sources.

Using the Fourier (frequency-)space decomposition convention given by
\begin{align}
	f[t] = \int_{\mathbb{R}} \frac{\dd\omega}{2\pi} e^{-i\omega t} \widetilde{f}[\omega] ,
\end{align}
I will also demonstrate, the corresponding frequency space expression of equations \eqref{Maxwell_DipoleEnergyRadiation_EvenDim} and \eqref{Maxwell_DipoleEnergyRadiation_OddDim} is
\begin{align}
	\label{Maxwell_DipoleEnergyRadiation_FreqSpace}
	\frac{\dd E}{\dd \omega}
	&= \frac{d-2}{2^{d+1} \pi^{\frac{d-1}{2}} \Gamma[\frac{d+1}{2}]} \omega^d \vec{d}[\omega]\cdot\vec{d}[\omega]^* ,
\end{align}
with $^*$ denoting complex conjugation. 

After deriving the closely related electromagnetic angular momentum flux, I will then proceed to examine the retarded history integral of a monochromatic dipole of duration $T$, which I view as a toy model of a finite duration light bulb.

In \S \eqref{Section_Gravity}, I will apply the experience gained in the electromagnetic calculations to the case of General Relativity (GR) with the cosmological constant set to zero. The $d=4$ gravitational quadrupole radiation formula
\begin{align}
\frac{\dd E}{\dd t}
= \frac{\GN}{5} \partial_t^3 Q_{ab}^{(\text{t})}[t-r] \partial_t^3 Q_{ab}^{(\text{t})}[t-r] ,
\end{align}
where $\GN$ is Newton's constant, is responsible for the decrease in the orbital period of the Hulse-Taylor binary pulsar system, a phenomenon verified to be consistent with GR at the sub-percent level \cite{Weisberg:2010zz}. This gave physicists confidence that gravitational waves do exist, long before their direct detection by the LIGO \cite{LIGO} and Virgo \cite{VIRGO} experiments. In even dimensions $d \geq 4$, I will find its generalization to read
\begin{align}
	\label{GR_EnergyFlux_EvenDim}
	\frac{\dd E}{\dd t}
	&= \frac{d(d-1)(d-3)\GN}{2^{d} (d-2) \pi^{\frac{d-5}{2}} \Gamma\left[\frac{d+3}{2}\right]}
	\partial_t^{\frac{d+2}{2}} Q^{(\text{t})}_{ab}[t-r] \partial_t^{\frac{d+2}{2}} Q^{(\text{t})}_{ab}[t-r] .
\end{align}
Here, the traceless quadrupole moment is defined in terms of the quadrupole $Q_{ab}$ as
\begin{align}
Q^{(\text{t})}_{ab}[t-r; \text{even $d \geq 4$}]
	&\equiv \left( \frac{1}{2} \delta_{a}^{\{i} \delta_{b}^{j\}} - \frac{\delta^{ij} \delta_{ab}}{d-1} \right) Q_{ij}[t-r] .
\end{align}
\footnote{In this paper, the convention for symmetrization and anti-symmetrization are: $S_{\{\alpha\beta\}} \equiv S_{\alpha\beta} + S_{\beta\alpha}$ and $A_{[\alpha\beta]} \equiv A_{\alpha\beta} - A_{\beta\alpha}$.}The $d-$dimensional Newton's constant is defined via Einstein's equations $G_{\mu\nu}=8\pi\GN T_{\mu\nu}$. On the other hand, in odd dimensions $d \geq 5$,\footnote{It is widely accepted that gravitational radiation does not exist in 3D. Below, I will provide a (partial) analysis that shows the far zone effective energy-momentum (pseudo-)tensor of gravitation is zero in (2+1)D.} I will demonstrate that
\begin{align}
	\label{GR_EnergyFlux_OddDim}
	\frac{\dd E}{\dd t}
	&= \frac{d (d-1) (d-3) \GN}{2^{d} (d-2) \pi^{\frac{d-3}{2}} \Gamma\left[\frac{d+3}{2}\right]}
	\partial_t^{\frac{d+3}{2}} Q^{(\text{t})}_{ab}[t-r] \partial_t^{\frac{d+3}{2}} Q^{(\text{t})}_{ab}[t-r] ;
\end{align}
where the dependence on the traceless quadrupole moments now features an additional integral over its entire past history due to the tail propagation of gravitational perturbations:
\begin{align}
Q^{(\text{t})}_{ab}[t-r; \text{odd $d \geq 5$}]
	&\equiv \left( \frac{1}{2} \delta_{a}^{\{i} \delta_{b}^{j\}} - \frac{\delta_{ab} \delta^{ij}}{d-1} \right)\int_{0}^{\infty} \frac{\dd\mu}{\mu^{\frac{1}{2}}} Q_{ij}[t-r-\mu] .	
\end{align}
Additionally, the angular frequency space counterpart to equations \eqref{GR_EnergyFlux_EvenDim} and \eqref{GR_EnergyFlux_OddDim} is\footnote{A word on notation: the $\GN$ here is the $G_d$ in eq. (4.25) of \cite{Cardoso:2008gn}. However, while our frequency space expressions are really energy or angular momentum per {\it angular} frequency ($\dd E/\dd \omega$ or $\dd L^{ij}/\dd \omega$) their answer is the energy loss per frequency ($\dd E/\dd \nu$), where $\omega[\text{here}] = 2\pi \nu$, even though they still call it $\dd E/\dd \omega$.}
\begin{align}
\label{GR_EnergyFlux_FreqSpace}
\frac{\dd E}{\dd \omega}
&= \frac{d (d-1) (d-3) \GN}{2^{d+1} (d-2) \pi^{\frac{d-3}{2}} \Gamma\left[\frac{d+3}{2}\right]}
\omega^{d+2} \widetilde{Q}^{(\text{t})}_{ab}[\omega] \widetilde{Q}^{(\text{t})}_{ab}[\omega]^* .
\end{align}
After working out the gravitational angular momentum flux -- which in 4D is responsible for circularizing initially eccentric binary systems -- I will then close in \S \eqref{Section_Byebye}. In \S \eqref{Section_Calculus} I discuss the solid angle tensor integrals used to obtain the total radiation rate ($\dd E/\dd t$ or $\dd L^{ij}/\dd t$) from the differential direction-dependent ones ($\dd E/(\dd t \dd \Omega)$ or $\dd L^{ij}/(\dd t \dd \Omega)$).

\section{Green's Functions in the Far Zone: $\omega r \to \infty$}
\label{Section_GreenFarZone}

Because the radiation formulas of equations \eqref{EnergyFlux} and \eqref{AngularMomentumFlux} involve the far zone $r\to\infty$ limits, the main objective of this section is to provide a step-by-step guide to lead the reader from the exact Green's functions in equations \eqref{G_EvenDim} and \eqref{G_OddDim} to their respective leading order $1/r^{(d/2)-1}$ and next-to-leading order $1/r^{d/2}$ far zone radiative limits in equations \eqref{G_EvenDim_FarZone} and \eqref{G_OddDim_FarZone} below. I shall then use the results to first solve explicitly the massless scalar wave equation in eq. \eqref{MasslessScalar_Solution}. As we will witness in the next two sections, the Lorenz gauge vector potential and the linear de Donder gauge gravitational perturbation can be directly obtained from eq. \eqref{MasslessScalar_Solution}. Since these solutions are already in the far zone $C_1/r^{(d/2)-1} + C_2/r^{d/2} + \dots$ form, the desired radiation formulas in equations \eqref{EnergyFlux} and \eqref{AngularMomentumFlux} then follow readily.

{\bf Driven SHO} \qquad First, we shall see that re-writing the Green's functions in equations \eqref{G_EvenDim} and \eqref{G_OddDim} in frequency space would allow us to perform a clean separation-of-variables, which will then facilitate this $1/r$ expansion.
\begin{align}
	\label{G_FrequencySpace}
	G_d[x-x']
	&= \int_{\mathbb{R}} \frac{\dd\omega}{2\pi} e^{-i\omega T} \widetilde{G}_d[\omega R] . \\
	T &\equiv t-t',
	\qquad
	R  \equiv |\vec{x}-\vec{x}'| .
\end{align}
Referring to eq. \eqref{MasslessScalar_Solution}, obtained by integrating $J$ against eq. \eqref{G_FrequencySpace} tells us $\omega$ corresponds to the angular frequency of the source producing these waves:
\begin{align}
	\label{DrivenSHO}
	\psi[t,\vec{x}] = \int_{\mathbb{R}} \frac{\dd\omega}{2\pi} e^{-i\omega t} \int_{\mathbb{R}^{d-1}} \dd^{d-1}\vec{x}' \widetilde{G}[\omega R] \widetilde{J}[\omega,\vec{x}'] ,
\end{align}
where $\widetilde{J}[\omega,\vec{x}'] = \int_{\mathbb{R}} \dd t' e^{i\omega t'} J[t',\vec{x}']$. The field $\psi$ in eq. \eqref{DrivenSHO} is simply the sum over harmonic oscillators, driven by $\widetilde{J}$; and analogous statements apply for the Lorenz gauge vector potential $A_\nu$ and the de Donder gauge gravitational perturbation $\bar{h}_{\mu\nu}$ just by replacing $\psi \to A_\nu$ and $J \to J_\nu$; or $\psi \to \bar{h}_{\mu\nu}$ and $J \to -16\pi \GN T_{\mu\nu}$.

{\bf Frequency Space and Separation-of-Variables} \qquad In even dimensions $d \geq 4$, we first employ the Fourier integral representation of the Dirac delta function
\begin{align}
	\delta[T-R] =  \int_{\mathbb{R}} \frac{\dd\omega}{2\pi} e^{-i \omega (T-R)}
\end{align}
on eq. \eqref{G_EvenDim}, followed by recalling that the Hankel function of the first kind with order one-half is
\begin{align}
	H^{(1)}_{\frac{1}{2}}[z] = -i \sqrt{\frac{2}{\pi z}} e^{iz} ,
\end{align}
to deduce
\begin{align}
	\widetilde{G}_{\text{even $d \geq 4$}}[\omega R]
	&= \frac{i \omega}{4\sqrt{2\pi}} \left( -\frac{1}{2\pi R} \frac{\partial}{\partial R} \right)^{\frac{d-4}{2}} \frac{H_{\frac{1}{2}}^{(1)}[\omega R]}{\sqrt{\omega R}} .
\end{align}
In odd dimensions $d \geq 3$, upon multiplying eq. \eqref{G_OddDim} by $e^{i\omega T}$ and integrating over $T \in \mathbb{R}$, we may first recognize the integral representation of the Hankel function
\begin{align}
	H_0^{(1)}[x>0] = -\frac{2i}{\pi} \int_{1}^{\infty} \frac{e^{ixt}}{\sqrt{t^2-1}} \dd t ,
\end{align}
followed by analytic continuation to all $x \in \mathbb{R}$, to infer
\begin{align}
	\widetilde{G}_{\text{odd $d \geq 3$}}[\omega R]
	&= \frac{i}{4} \left( -\frac{1}{2\pi R} \frac{\partial}{\partial R} \right)^{\frac{d-3}{2}} H_{0}^{(1)}[\omega R] .
\end{align}
Finally, let us utilize the identity, for non-negative integers $n=0,1,2,3,\dots$,
\begin{align}
	\left( \frac{1}{z} \frac{\dd}{\dd z} \right)^n \frac{H_{\nu}^{(1)}[z]}{z^\nu}
	= (-)^n \frac{H_{\nu+n}^{(1)}[z]}{z^{\nu+n}}
\end{align}
to arrive at the following frequency space Green's functions for all $d \geq 3$.
\begin{align}
	\label{G_EvenDim_FreqSpace}
	\widetilde{G}_{d = 4+2n}[\omega R]
	&= \frac{i \omega^{2n+1}}{4(2\pi)^{\frac{1}{2}+n}} \frac{H_{\frac{1}{2}+n}^{(1)}[\omega R]}{(\omega R)^{\frac{1}{2}+n}} \\
	\label{G_OddDim_FreqSpace}
	\widetilde{G}_{d = 3+2n}[\omega R]
	&= \frac{i \omega^{2n}}{4(2\pi)^{n}} \frac{H_{n}^{(1)}[\omega R]}{(\omega R)^{n}}
\end{align}
The factor $H_\nu^{(1)}[ \omega R]/(\omega R)^\nu$ obeys addition formulas that separates the $r \equiv |\vec{x}|$ and $r' \equiv |\vec{x}'|$ dependence in $R = |\vec{x}-\vec{x}'|$. Denoting $r_< \equiv \min[r,r']$, $r_> \equiv \max[r,r']$, $\widehat{r} \equiv \vec{x}/r$ and $\widehat{r}' \equiv \vec{x}'/r'$,
\begin{align}
	H_0^{(1)}[\omega R] 
	&= \sum_{\ell=-\infty}^{+\infty} J_\ell[\omega r_<] H_\ell^{(1)}[\omega r_>] e^{i\ell\phi}, \\
	\frac{H_\nu^{(1)}[\omega R]}{(\omega R)^\nu}
	&= 2^\nu \Gamma[\nu] \sum_{\ell=0}^{+\infty} (\nu+\ell) \frac{J_{\nu+\ell}[\omega r_<]}{(\omega r_<)^\nu} \frac{H_{\nu+\ell}^{(1)}[\omega r_>]}{(\omega r_>)^\nu} C_\ell^{(\nu)}[\widehat{r} \cdot \widehat{r}'],
	\qquad
	\nu \neq 0,-1,-2,-3,\dots .
\end{align}
For all even dimensions $d = 4+2n \geq 4$, therefore,
\begin{align}
	\label{G_EvenDim_SeparatedForm}
	\widetilde{G}_{4+2n}[\omega R]
	&= \frac{i \omega^{1+2n}}{4(2\pi)^{\frac{1}{2}+n}} 2^{\frac{1}{2}+n} \Gamma\left[ \frac{1}{2}+n \right] \nonumber\\
	&\qquad \times
	\sum_{\ell=0}^{+\infty} \left( \ell + \frac{1}{2} + n \right) \frac{J_{\frac{1}{2}+n+\ell}[\omega r_<]}{(\omega r_<)^{\frac{1}{2}+n}} \frac{H_{\frac{1}{2}+n+\ell}^{(1)}[\omega r_>]}{(\omega r_>)^{\frac{1}{2}+n}} C_\ell^{\left(\frac{1}{2}+n\right)}\left[ \widehat{r} \cdot \widehat{r}' \right] ,
	\qquad
	n=0,1,2,3,\dots ;
\end{align}
where $J_\nu[z]$ is the Bessel function, $C_\ell^{(\nu)}[z]$ is Gegenbauer’s polynomial. (For the 4D case, recognizing $C_\ell^{(\frac{1}{2})}$ to be $P_\ell$, the Legendre polynomial, would recover the familiar result found in most advanced electromagnetism textbooks.) And for odd dimensions $d = 3+2n \geq 3$,
\begin{align}
	\label{G_OddDim_SeparatedForm_3D}
	\widetilde{G}^+_{3}[\omega R]
	&= \frac{i}{4} \sum_{\ell=-\infty}^{+\infty} J_\ell[\omega r_<] H_\ell^{(1)}[\omega r_>] e^{i\ell \phi} ,
	\qquad \widehat{r} \cdot \widehat{r}' \equiv \cos\phi , \\
	\label{G_OddDim_SeparatedForm_dgeq5}
	\widetilde{G}^+_{3+2n}[\omega R]
	&= \frac{i \omega^{2n}}{4 (2\pi)^n} 2^n \Gamma[n] \sum_{\ell=0}^{+\infty} (n+\ell) \frac{J_{n+\ell}[\omega r_<]}{(\omega r_<)^n} \frac{H_{n+\ell}^{(1)}[\omega r_>]}{(\omega r_>)^n} C_\ell^{(n)}\left[ \widehat{r} \cdot \widehat{r}' \right], \qquad n = 1,2,3,\dots .
\end{align}
{\bf Far Zone: Frequency Space} \qquad For our radiation calculations, $r$ the observer-source distance is always much larger than $r'$, which is at most the size of the source itself, since we will be integrating $\vec{x}'$ against the electromagnetic current or the stress-energy tensor of matter. (Recall: we will always place $\vec{0}$ inside the source.) The $\omega r$ dependence therefore occurs in the factor $H_\nu^{(1)}[\omega r]/(\omega r)^\nu$ in equations \eqref{G_EvenDim_SeparatedForm} through \eqref{G_OddDim_SeparatedForm_dgeq5}. If we then replace these $H_\nu^{(1)}[\omega r]$ with their large argument expansions -- a finite power series for $\nu = \frac{1}{2} + \ell + n$ (even dimensions) and an asymptotic one for $\nu = n+\ell$ (odd dimensions) --
\begin{align}
	H_\nu^{(1)}[\omega r]
	&=  \frac{2}{\sqrt{2 \pi \omega r}} e^{i(\omega r - \frac{\pi}{2} \nu - \frac{\pi}{4})} \left( 1
	+ \frac{i}{2} \frac{\left( \nu - \frac{1}{2} \right)\left( \nu + \frac{1}{2} \right)}{\omega r}
	+ \mathcal{O}[(\omega r)^{-2}] \right) ,
\end{align}
the even dimensional result in eq. \eqref{G_EvenDim_SeparatedForm} may now evaluated in the far zone $\omega r \to \infty$ as
\begin{align}
	\label{G_EvenDim_SeparatedForm_FarZone_v1}
	&\widetilde{G}_{4+2n \geq 4}[\omega R]
	= \frac{(-i\omega)^{n}}{2(2\pi r)^{1+n}} 2^{\frac{1}{2}+n} \Gamma\left[ \frac{1}{2}+n \right] e^{i \omega r} \\
	&\qquad
	\times \sum_{\ell=0}^{+\infty} (-i)^\ell
	\left( \ell + \frac{1}{2} + n \right) \frac{J_{\frac{1}{2}+n+\ell}[\omega r']}{(\omega r')^{\frac{1}{2}+n}}
	\left( 1 + \frac{i}{2} \frac{n(n+1) + \ell(\ell+2n+1)}{\omega r} + \mathcal{O}[(\omega r)^{-2}]\right)
	C_\ell^{\left(\frac{1}{2}+n\right)}\left[ \widehat{r} \cdot \widehat{r}' \right] . \nonumber
\end{align}
Whereas the same $\omega r \to \infty$ far zone limit of the odd dimensional results in eq. \eqref{G_OddDim_SeparatedForm_3D}, with $\widehat{r} \cdot \widehat{r}' \equiv \cos\phi$, becomes
\begin{align}
	\label{G_OddDim_SeparatedForm_3D_FarZone_v1}
	\widetilde{G}_{3}[\omega R]
	&= \frac{i}{2 \sqrt{2\pi \omega r}} e^{i (\omega r - \frac{\pi}{4})}  \sum_{\ell=-\infty}^{+\infty} (-i)^\ell J_\ell[\omega r']
	\left( 1
	+ \frac{i}{2} \left(\frac{- \frac{1}{4} + \ell^2}{\omega r}\right)
	+ \mathcal{O}[(\omega r)^{-2}] \right) e^{i\ell \phi} ;
\end{align}
and that in eq. \eqref{G_OddDim_SeparatedForm_dgeq5} turns into
\begin{align}
	\label{G_OddDim_SeparatedForm_dgeq5_FarZone_v1}
	&\widetilde{G}_{3+2n \geq 5}[\omega R]
	= \frac{(-i)^{n-1} \omega^{2n} (n-1)!}{4 \pi^n \sqrt{2\pi} (\omega r)^{\frac{1}{2}+n}} e^{i (\omega r - \frac{\pi}{4})} \\
	&\qquad\times
	\sum_{\ell=0}^{+\infty} (-i)^{\ell} (2n+2\ell) \frac{J_{n+\ell}[\omega r']}{(\omega r')^n} \left( 1
	+ \frac{i}{2} \frac{n^2 - \frac{1}{4} + \ell(\ell+2n)}{\omega r}
	+ \mathcal{O}[(\omega r)^{-2}] \right) C_\ell^{(n)}\left[ \widehat{r} \cdot \widehat{r}' \right]  . \nonumber
\end{align}
Next, we recognize the $\ell(\ell+2n+1)$, $\ell^2$, and $\ell(\ell+2n)$ occurring within the summations in equations \eqref{G_EvenDim_SeparatedForm_FarZone_v1} through \eqref{G_OddDim_SeparatedForm_dgeq5_FarZone_v1} as the eigenvalue $\ell(\ell+d-3)$ of the negative Laplacian on the $(d-2)-$sphere, for all $d \geq 3$. Specifically, we may replace them with the negative Laplacian acting on the $e^{i\ell\phi}$ or Gegenbauer polynomial $C_\ell^{(\frac{d-3}{2})}$ because
\begin{align}
	-\vec{\nabla}_{\mathbb{S}^{1}}^2 e^{i\ell \phi}
	&= \ell^2 e^{i\ell \phi}, \qquad (d=3) ; \\
	-\vec{\nabla}_{\mathbb{S}^{d-2}}^2 C_\ell^{\left(\frac{d-3}{2}\right)}[\widehat{r} \cdot \widehat{r}']
	&= \ell(\ell + d-3) C_\ell^{\left(\frac{d-3}{2}\right)}[\widehat{r} \cdot \widehat{r}'] ,
	\qquad (d \geq 4) .
\end{align}
Upon the replacement $\ell(\ell+d-3) \to -\vec{\nabla}_{\mathbb{S}^{d-2}}^2$ in equations \eqref{G_EvenDim_SeparatedForm_FarZone_v1} through \eqref{G_OddDim_SeparatedForm_dgeq5_FarZone_v1}, we will recognize the remaining summations to be nothing but the Bessel function expansion of the plane wave. In $d-1 = 2$ spatial dimensions,
\begin{align}
	e^{i\vec{k}\cdot\vec{x}}
	&= \sum_{\ell=-\infty}^{+\infty} i^\ell J_\ell[kr] e^{i\ell\phi} ;
\end{align}
and in three and higher spatial dimensions, $d-1 \geq 3$,
\begin{align}
	e^{i\vec{k}\cdot\vec{x}}
	&= 2^{\frac{d-3}{2}} \Gamma\left[\frac{d-3}{2}\right] \sum_{\ell=0}^{\infty} \left(\frac{d-3}{2}+\ell\right) i^\ell \frac{J_{\frac{d-3}{2}+\ell}[k r]}{(k r)^{\frac{d-3}{2}}} C_\ell^{\left( \frac{d-3}{2} \right)}[\widehat{r} \cdot \widehat{r}'] .
\end{align}
{\it Results} \qquad We have arrived at the far zone $\omega r \to \infty$ frequency space Green's functions. The even $(d \geq 4)$ and odd $(d \geq 3)$ dimensional Green's functions are, respectively,
{\allowdisplaybreaks\begin{align}
		\label{G_EvenDim_FarZone_v2}
		\widetilde{G}_{4+2n \geq 4}[\omega R]
		&= \frac{(-i \omega)^n}{2(2\pi r)^{1+n}} \left( 1 + \frac{i}{2} \frac{n(n+1) - \vec{\nabla}_{\mathbb{S}^{2n+2}}^2}{\omega r} + \mathcal{O}[ (\omega r)^{-2}] \right) e^{i \omega (r - \widehat{r} \cdot \vec{x}')} , \\
		\label{G_OddDim_FarZone_v2}
		\widetilde{G}_{3+2n \geq 3}[\omega R]
		&= \frac{(-i \omega)^n}{2 (2\pi r)^n \sqrt{2\pi} \sqrt{-i\omega r}}
		\left( 1
		+ \frac{n^2 - \frac{1}{4} - \vec{\nabla}_{\mathbb{S}^{2n+1}}^2}{2 (-i \omega r)}
		+ \mathcal{O}[(\omega r)^{-2}] \right) e^{i\omega(r - \widehat{r} \cdot \vec{x}')} .
\end{align}}
To carry out the derivatives associated with $\vec{\nabla}_{\mathbb{S}^{d-2}}^2$, let us record that: the Laplacian on $\mathbb{S}^{d-2}$ acting on a function that depends on angles solely through the object $c \equiv \widehat{r} \cdot \widehat{r}'$ is, for all $d \geq 3$,
\begin{align}
	\vec{\nabla}^2_{\mathbb{S}^{d-2}} \psi\left[ \widehat{r} \cdot \widehat{r}' \right]
	&= \frac{1}{(1-c^2)^{\frac{d-4}{2}}} \partial_c \left( (1-c^2)^{\frac{d-2}{2}} \partial_c \psi\left[ \widehat{r} \cdot \widehat{r}' \right] \right) \\
	\label{LaplacianOnSphere}
	&= (1-c^2) \psi''[c] - (d-2)c \psi'[c] .
\end{align}
The expanded forms of equations \eqref{G_EvenDim_FarZone_v2} and \eqref{G_OddDim_FarZone_v2} then read
{\allowdisplaybreaks\begin{align}
		\label{G_EvenDim_FarZone_v3}
		\widetilde{G}_{4+2n \geq 4}&[\omega R]
		= \frac{(-i \omega)^n}{2(2\pi r)^{1+n}} \\
		&\times
		\left( 1 + \frac{1}{2} \frac{n(n+1) + (2n+2) (-i\omega) (\widehat{r}\cdot\vec{x}') - (-i\omega)^2 (r'^2 - (\widehat{r}\cdot\vec{x}')^2)}{-i\omega r} + \mathcal{O}[ (\omega r)^{-2}] \right) e^{i \omega (r - \widehat{r} \cdot \vec{x}')} , \nonumber\\
		\label{G_OddDim_FarZone_v3}
		\widetilde{G}_{3+2n \geq 3}&[\omega R]
		= \frac{(-i \omega)^n}{2 (2\pi r)^{n+\frac{1}{2}} \sqrt{-i\omega}} \\
		&\times
		\left( 1
		+ \frac{1}{2} \frac{n^2 - \frac{1}{4} + (2n+1) (-i\omega) (\widehat{r}\cdot\vec{x}') - (-i\omega)^2 (r'^2 - (\widehat{r}\cdot\vec{x}')^2)}{(-i \omega r)}
		+ \mathcal{O}[(\omega r)^{-2}] \right) e^{i\omega(r - \widehat{r} \cdot \vec{x}')} . \nonumber
\end{align}}
{\it Relativistic corrections} \qquad Before moving on, I wish to highlight the presence of the $-\widehat{r} \cdot \vec{x}'$ in the exponential $e^{i(\omega r - \widehat{r}\cdot\vec{x}')}$ as a relativistic correction. By examining the $e^{-i\omega T} \widetilde{G}$ in eq. \eqref{G_FrequencySpace}, we see that the combination $e^{-i\omega(t-t'-r)}$ arising from the expressions in equations \eqref{G_EvenDim_FarZone_v3} and \eqref{G_OddDim_FarZone_v3} describes an outgoing spherical wave, with angular frequency $\omega$ associated with that of the source. Since $-\widehat{r}\cdot\vec{x}'$ scales as the characteristic size of the source $r_s$, it does not produce an appreciable phase shift as long as $\omega \cdot (\widehat{r} \cdot \vec{x}') \equiv (2\pi f)(\widehat{r} \cdot \vec{x}')$ is much less than $2\pi$. Physically, this indicates: as long as the characteristic timescale of the source ($t_s \sim 1/f$) is much slower than its characteristic size -- namely, $\omega r_s \sim 2\pi (r_s/t_s) \ll 2\pi$ -- then this factor is negligible. To further corroborate this relativistic correction interpretation, also observe that $r_s$ is, in natural $c=1$ units, the light-crossing time of the source; i.e., the non-relativistic limit is simply the situation where the light-crossing time is much shorter than the characteristic time scale of the source itself.

{\bf Far Zone: Real-time} \qquad The real-time far zone radiative Green's function requires that we perform the Fourier integral in eq. \eqref{G_FrequencySpace}. To this end, we recognize all positive powers of $-i \omega$ to be time derivatives: namely, $(-i \omega)^n e^{-i\omega T} = \partial_t^n e^{-i\omega T}$. Note that the $n(n+1)/(-i\omega r)$ term in eq. \eqref{G_EvenDim_FarZone_v3} is non-zero only for $n \geq 1$, so together with the $(-i \omega)^n$ pre-factor, we see that it contains $n-1$ time derivatives for $d=4+2n > 4$. We then arrive at the far zone (radiation) Green's function in even dimensions $d = 4+2n \geq 4$:
\begin{align}
	\label{G_EvenDim_FarZone}
	G_{4+2n}[x-x']
	&= \frac{1}{2(2\pi r)^{1+n}} \Bigg( \partial_t^n
	+ \frac{1}{2} \frac{n(n+1)}{r} \partial_t^{n-1} \\
	&\qquad\qquad
	+ \frac{1}{2} \frac{(\widehat{r} \cdot \vec{x}') (2n+2) - (\vec{x}'^2 - (\widehat{r} \cdot \vec{x}')^2) \partial_t}{r} \partial_t^{n}
	+ \mathcal{O}[r^{-2}] \Bigg) \delta[t-t'-r+\widehat{r} \cdot \vec{x}'] . \nonumber
\end{align}
The odd dimensional case in eq. \eqref{G_OddDim_FarZone_v3} requires the following manipulation due to the presence of the inverse fractional powers of frequencies, $1/(-i\omega )^{\frac{1}{2}}$ at order $1/r^{\frac{1}{2}+n}$ and $1/(-i\omega)^{\frac{3}{2}}$ at order $1/r^{\frac{3}{2}+n}$. By invoking the representation of the Gamma function -- for Re$[z]>0$ and Im$[\alpha]>0$ --
\begin{align}
\label{GammaFunction}
	\frac{1}{(-i \alpha)^z}
	= \frac{1}{\Gamma[z]} \int_{0}^{\infty} \dd\mu \mu^{z-1} \exp\left[ i \mu \cdot \alpha \right] ,
\end{align}
where $z=\frac{1}{2},\frac{3}{2},\dots$; and replacing $\alpha \rightarrow \omega + i0^+$, eq. \eqref{G_OddDim_FarZone_v3} is transformed into
\begin{align}
	\label{G_OddDim_FarZone_v4}
	\widetilde{G}_{3+2n}[\omega R]
	= \frac{(-i \omega)^n}{2 \sqrt{\pi} (2\pi r)^{n+\frac{1}{2}}} \int_{0}^{\infty} \dd\mu
	e^{- \mu \cdot 0^+}
	\left( \mu^{-\frac{1}{2}}
	+ \mu^{\frac{1}{2}} \frac{n^2 - \frac{1}{4} - \vec{\nabla}_{\mathbb{S}^{2n+1}}^2}{r}
	+ \mathcal{O}[r^{-1}] \right) e^{i\omega (r-\widehat{r}\cdot\vec{x}' + \mu)} .
\end{align}
Here, we have replaced $(2n+1) (-i\omega) (\widehat{r}\cdot\vec{x}') - (-i\omega)^2 (r'^2 - (\widehat{r}\cdot\vec{x}')^2)$ with $-\vec{\nabla}_{\mathbb{S}^{2n+1}}^2$ for compactness of notation. Multiplying eq. \eqref{G_OddDim_FarZone_v4} by $e^{-i\omega T}$, replacing $(-i\omega)^n \to \partial_t^n$, and integrating over $T \in \mathbb{R}$ hands us the far zone (radiation) Green's function in odd dimensions $d = 3+2n \geq 3$:
{\allowdisplaybreaks\begin{align}
		\label{G_OddDim_FarZone}
		&G_{3+2n}[x-x']
		= \frac{1}{\sqrt{2} (2\pi)^{n+1} \cdot r^{n + \frac{1}{2}}} \partial_t^n \int_{0}^{\infty} \dd\mu \exp\left[ - \mu \cdot 0^+ \right] \\
		&\qquad \times \left( \mu^{-\frac{1}{2}}
		+ \mu^{\frac{1}{2}} \frac{n^2 - \frac{1}{4} + (\widehat{r} \cdot \vec{x}') (2n+1) \partial_t - (r'^2-(\widehat{r}\cdot\vec{x}')^2)\partial_t^2}{r}
		+ \mathcal{O}[r^{-2}] \right)
		\delta\left[ t -t' - r - \mu + \widehat{r}\cdot\vec{x}' \right] . \nonumber
\end{align}}
{\it Massless Scalar in Even Dimensions} \qquad Plugging eq. \eqref{G_EvenDim_FarZone} into eq. \eqref{MasslessScalar_Solution} tells us the far zone massless scalar solution in even $d=4+2n$ takes the form
{\allowdisplaybreaks\begin{align}
		\label{MasslessScalar_EvenDim}
		\psi[t,\vec{x}]
		&= \frac{1}{2(2\pi r)^{1+n}} \int_{\mathbb{R}^{3+2n}} \dd^{3+2n}\vec{x}' \Bigg( \partial_t^n J[t-r+\widehat{r} \cdot \vec{x}',\vec{x}'] \\
		&\qquad\qquad
		+ \frac{1}{2} \frac{n(n+1) \partial_t^{n-1} + (\widehat{r} \cdot \vec{x}') (2n+2) \partial_t^n - (r'^2-(\widehat{r} \cdot \vec{x}')^2) \partial_t^{n+1}}{r} J[t-r+\widehat{r} \cdot \vec{x}',\vec{x}'] + \mathcal{O}[r^{-2}] \Bigg) ; \nonumber
\end{align}}
and its first and second derivatives are
{\allowdisplaybreaks\begin{align}
		\label{MasslessScalar_PD1_EvenDim}
		&\partial_\alpha \psi[t,\vec{x}] \\
		&= \frac{1}{2(2\pi r)^{1+n}} \int_{\mathbb{R}^{3+2n}} \dd^{3+2n}\vec{x}' \Bigg( \left(\delta_\alpha^0 - \delta_\alpha^l \widehat{r}^l\right) \partial_t^{n+1} J[t-r+\widehat{r} \cdot \vec{x}',\vec{x}'] \nonumber \\
		&\qquad
		+ \delta_\alpha^a P^{ab} \frac{x'^b}{r} \partial_t^{n+1} J[t-r+\widehat{r} \cdot \vec{x}',\vec{x}']
		-\frac{n+1}{r} \delta_\alpha^l \widehat{r}^l \partial_t^n J[t-r+\widehat{r} \cdot \vec{x}',\vec{x}'] \nonumber\\
		&\qquad
		+ \frac{1}{2} \frac{n(n+1) \partial_t^{n-1} + (\widehat{r} \cdot \vec{x}') (2n+2) \partial_t^n - (r'^2-(\widehat{r} \cdot \vec{x}')^2) \partial_t^{n+1}}{r} \left(\delta_\alpha^0 - \delta_\alpha^l \widehat{r}^l\right) J[t-r+\widehat{r} \cdot \vec{x}',\vec{x}'] + \mathcal{O}[r^{-2}] \Bigg)  \nonumber
\end{align}}
and
{\allowdisplaybreaks\begin{align}
		\label{MasslessScalar_PD2_EvenDim}
		&\partial_\alpha \partial_\beta \psi[t,\vec{x}] \\
		&= \frac{1}{2(2\pi r)^{1+n}} \int_{\mathbb{R}^{3+2n}} \dd^{3+2n}\vec{x}' \Bigg( \left(\delta_\alpha^0 - \delta_\alpha^l \widehat{r}^l\right) \left(\delta_\beta^0 - \delta_\beta^k \widehat{r}^k \right) \partial_t^{n+2} J[t-r+\widehat{r} \cdot \vec{x}',\vec{x}'] 
		- \delta_\alpha^l \frac{P^{lk}}{r} \delta_\beta^k \partial_t^{n+1} J[t-r+\widehat{r} \cdot \vec{x}',\vec{x}'] \nonumber \\
		&\qquad
		+ \delta_{\{\alpha}^a P^{ab} \frac{x'^b}{r} \left(\delta_{\beta\}}^0 - \delta_{\beta\}}^k \widehat{r}^k \right) \partial_t^{n+2} J[t-r+\widehat{r} \cdot \vec{x}',\vec{x}']
		-\frac{n+1}{r} \delta_{\{\alpha}^l \widehat{r}^l \left(\delta_{\beta\}}^0 - \delta_{\beta\}}^k \widehat{r}^k \right) \partial_t^{n+1} J[t-r+\widehat{r} \cdot \vec{x}',\vec{x}'] \nonumber\\
		&\qquad
		+ \frac{1}{2} \frac{n(n+1) \partial_t^{n} + (\widehat{r} \cdot \vec{x}') (2n+2) \partial_t^{n+1} - (r'^2-(\widehat{r} \cdot \vec{x}')^2) \partial_t^{n+2}}{r} \left(\delta_\alpha^0 - \delta_\alpha^l \widehat{r}^l\right) \left(\delta_\beta^0 - \delta_\beta^k \widehat{r}^k \right) J[t-r+\widehat{r} \cdot \vec{x}',\vec{x}'] + \mathcal{O}[r^{-2}] \Bigg) . \nonumber
\end{align}}
We have defined
\begin{align}
	\label{Pab}
	P^{ab} &\equiv \delta^{ab} - \widehat{r}^a \widehat{r}^b ,
\end{align}
which is orthogonal to the unit radial vector $\widehat{r}$ and also acts as a projector,
\begin{align}
	\label{Pab_Properties}
	\widehat{r}^a P_{ab} = 0
	\qquad \text{ and } \qquad
	P_{ab} P_{bc} = P_{ac} .
\end{align}
{\it Massless Scalar in Odd Dimensions} \qquad Along similar lines as the even dimensional case, plugging eq. \eqref{G_OddDim_FarZone} into eq. \eqref{MasslessScalar_Solution} tells us the far zone massless scalar solution in odd $d=3+2n$ takes the form
\begin{align}
	\label{MasslessScalar_OddDim}
	\psi[t,\vec{x}]
	&= \frac{1}{\sqrt{2} (2\pi)^{n+1} \cdot r^{n+\frac{1}{2}}} \int_{\mathbb{R}^{2+2n}} \dd^{2+2n}\vec{x}' \int_{0}^{\infty} \dd\mu \exp\left[ - \mu \cdot 0^+ \right]
	\Bigg( \mu^{-\frac{1}{2}} \partial_\tau^n J\left[ \tau,\vec{x}' \right] \\
	&+ \frac{\mu^{\frac{1}{2}}}{r} \left( \left(n^2 - \frac{1}{4}\right) \partial_\tau^n J\left[ \tau,\vec{x}' \right] + (\widehat{r} \cdot \vec{x}') (2n+1) \partial_\tau^{n+1} J\left[ \tau,\vec{x}' \right] - (r'^2 - (\widehat{r}\cdot\vec{x}')^2) \partial_\tau^{n+2} J\left[ \tau,\vec{x}' \right] \right)
	+ \mathcal{O}[r^{-2}] \Bigg) , \nonumber\\
	\tau &\equiv t - r - \mu + \widehat{r}\cdot\vec{x}' ; \nonumber
\end{align}
and its first derivative is
{\allowdisplaybreaks\begin{align}
		\label{MasslessScalar_PD1_OddDim}
		&\partial_\alpha \psi[t,\vec{x}] \nonumber\\
		&= \frac{1}{\sqrt{2} (2\pi)^{n+1} \cdot r^{n+\frac{1}{2}}} \int_{\mathbb{R}^{2+2n}} \dd^{2+2n}\vec{x}' \int_{0}^{\infty} \dd\mu \exp\left[ - \mu \cdot 0^+ \right] \\
		&\times 
		\Bigg\{ \mu^{-\frac{1}{2}} \left( \delta_\alpha^0 - \delta_\alpha^j \widehat{r}^j \right) \partial_\tau^{n+1} J\left[ \tau,\vec{x}' \right] + \frac{\mu^{-\frac{1}{2}}}{r} \left( \delta_\alpha^{a} P^{ab} x'^b \partial_\tau^{n+1} J\left[ \tau,\vec{x}' \right]  - \left( n + \frac{1}{2} \right) \widehat{r}^l \delta_\alpha^l \partial_\tau^n J\left[ \tau,\vec{x}' \right] \right) \nonumber\\
		&+ \frac{\mu^{-\frac{1}{2}}}{2r} \left( \left(n^2 - \frac{1}{4}\right) \partial_\tau^{n} J\left[ \tau,\vec{x}' \right] + (\widehat{r} \cdot \vec{x}') (2n+1) \partial_\tau^{n+1} J\left[ \tau,\vec{x}' \right] - (r'^2  - (\widehat{r}\cdot\vec{x}')^2) \partial_\tau^{n+2} J\left[ \tau,\vec{x}' \right] \right) \left( \delta_\alpha^0 - \delta_\alpha^j \widehat{r}^j \right) 
		+ \mathcal{O}[r^{-2}] \Bigg\} . \nonumber
\end{align}}
In the last line of eq. \eqref{MasslessScalar_PD1_OddDim}, we have converted one of the $\tau$ derivatives into a negative $\mu$ derivative (i.e., $\partial/\partial\tau = -\partial/\partial \mu$), and integrated it by parts. The surface term at $\mu=\infty$ is zero because of $e^{-\mu \cdot 0^+}$ and that at $\mu=0$ is zero because of $\mu^{1/2}$.

Finally, the second derivative of eq. \eqref{MasslessScalar_OddDim} is
{\allowdisplaybreaks\begin{align}
		\label{MasslessScalar_PD2_OddDim}
		\partial_\alpha \partial_\beta \psi[t,\vec{x}] 
		&= \frac{1}{\sqrt{2} (2\pi)^{n+1} \cdot r^{n+\frac{1}{2}}} \int_{\mathbb{R}^{2+2n}} \dd^{2+2n}\vec{x}' \int_{0}^{\infty} \dd\mu \exp\left[ - \mu \cdot 0^+ \right] \\
		&\times 
		\Bigg\{ \mu^{-\frac{1}{2}} \left( \delta_\alpha^0 - \delta_\alpha^j \widehat{r}^j \right) \left( \delta_\beta^0 - \delta_\beta^k \widehat{r}^k \right) \partial_\tau^{n+2} J\left[ \tau,\vec{x}' \right]
		- \frac{\mu^{-\frac{1}{2}}}{r} \delta_\alpha^a P^{ab} \delta^b_\beta \partial_\tau^{n+1} J[\tau,\vec{x}'] \nonumber\\
		&\qquad\qquad+ \frac{\mu^{-\frac{1}{2}}}{r} \left( \delta_{\{\alpha}^{a} P^{ab} x'^b \partial_\tau^{n+2} J\left[ \tau,\vec{x}' \right]  - \left( n + \frac{1}{2} \right) \widehat{r}^l \delta_{\{\alpha}^l \partial_\tau^{n+1} J\left[ \tau,\vec{x}' \right] \right) \left( \delta_{\beta\}}^0 - \delta_{\beta\}}^k \widehat{r}^k \right) \nonumber\\
		&\qquad\qquad+ \frac{\mu^{-\frac{1}{2}}}{2r} \Bigg( \left(n^2 - \frac{1}{4}\right) \partial_\tau^{n+1} J\left[ \tau,\vec{x}' \right] + (\widehat{r} \cdot \vec{x}') (2n+1) \partial_\tau^{n+2} J\left[ \tau,\vec{x}' \right] \nonumber\\
		&\qquad\qquad\qquad\qquad
		- (r'^2-(\widehat{r}\cdot\vec{x}')^2) \partial_\tau^{n+3} J\left[ \tau,\vec{x}' \right] \Bigg) \left( \delta_\alpha^0 - \delta_\alpha^j \widehat{r}^j \right)  \left( \delta_\beta^0 - \delta_\beta^k \widehat{r}^k \right) 
		+ \mathcal{O}[r^{-2}] \Bigg\} . \nonumber
\end{align}}
We now turn to tackle Maxwell's electromagnetism and the weak field and zero cosmological constant limits of Einstein's General Relativity.

\section{Maxwell's Electromagnetism}
\label{Section_EM}
{\bf Setup} \qquad Maxwell's electromagnetism, sourced by a conserved current $J^\nu$, is defined by
\begin{align}
\label{Maxwell_Eqn}
\partial_\mu F^{\mu\nu} = J^\nu
\qquad \text{and} \qquad
\partial_\nu J^\nu = 0 .
\end{align}
The anti-symmetric electromagnetic tensor $F_{\mu\nu} = -F_{\nu\mu}$ itself is built out of the vector potential $A_\mu$,
\begin{align}
F_{\mu\nu}
\equiv \partial_{[\mu} A_{\nu]}
\equiv \partial_{\mu} A_{\nu} - \partial_{\nu} A_{\mu} .
\end{align}
Given an inertial frame, the $d-1$ components of $F_{0i} = -F_{i0}$ are electric fields; and the $(1/2)(d-1)(d-2)$ components of $F_{ij} = -F_{ji}$ are the magnetic fields. (The diagonal terms of $F_{\mu\nu}$ are zero.) The electric field may always be regarded as a spatial vector, but only in $d=4$ can the magnetic ones be interpreted as such -- i.e., it is the only admissible solution to $(1/2)(d-1)(d-2) = d-1$.

{\bf Energy Flux} \qquad The conserved and symmetric energy-momentum tensor of the electromagnetic fields is
\begin{align}
\label{Maxwell_StressTensor}
T^{\mu\nu}
= -F^{\mu\sigma} F^\nu_{\phantom{\nu}\sigma} + \frac{1}{4} \eta^{\mu\nu} F^{\rho\sigma} F_{\rho\sigma} .
\end{align}
In particular, the momentum density responsible for carrying energy to infinity is
\begin{align}
T^{0i} = -F_{0j} F_{ij} .
\end{align}
Because the anti-symmetric magnetic field $F_{ij}$ does not exist in $d=2$ dimensions, we shall only focus on $d \geq 3$ in this section. Note that conservation $\partial_\mu T^{\mu\nu} = 0$ is not an identity, but only holds when the energy-momentum is evaluated on the solutions to Maxwell's equations in \eqref{Maxwell_Eqn}.

In terms of the electromagnetic fields and the outward point unit radial vector $\widehat{r}$, the energy flux in eq. \eqref{EnergyFlux} is
\begin{align}
\frac{\dd E}{\dd t \dd\Omega}
\label{Maxwell_EnergyFlux}
&= -\lim_{r \to \infty} r^{d-2} F_{0j} F_{ij} \widehat{r}^i .
\end{align}
{\bf Angular Momentum Flux} \qquad For a fixed and distinct pair $(i,j)$, the angular momentum current is defined as
\begin{align}
	J^{ij \mu} = x^{[i} T^{j] \mu} .
\end{align}
In 4D, the $J^{ij 0} = x^{[i} T^{j]0}$ is simply the spatial Hodge dual of the perhaps more familiar expression involving the cross product between the position vector and the momentum density; namely, $L^i = (1/2) \epsilon^{ijk} J^{jk0} = \epsilon^{ijk} x^j T^{k0}$, where $\epsilon^{ijk}$ is the Levi-Civita symbol with $\epsilon^{123} \equiv 1$. One may also readily check that this current is conserved, $\partial_\mu J^{ij \mu} = 0$, due to the symmetry of the stress tensor ($T^{\mu\nu} = T^{\nu\mu}$) and its `on-shell' conservation ($\partial_\mu T^{\mu\nu} = 0$). The angular momentum flux in eq. \eqref{AngularMomentumFlux} is thus
\begin{align}
\label{Maxwell_AngularMomentumFlux}
\frac{\dd L^{ij}}{\dd t \dd\Omega}
&= \lim_{r \to \infty} r^{d-2} x^{[i} T^{j]k} \widehat{r}^k .
\end{align}
The primary goal of this section is to compute equations \eqref{Maxwell_EnergyFlux} and \eqref{Maxwell_AngularMomentumFlux} for Maxwell's electromagnetism. Because they involve the $r \to \infty$ limit, we do not need the exact $F_{\mu\nu}$, but only its asymptotic far zone expression developed in powers of $1/r$: for eq. \eqref{Maxwell_EnergyFlux}, to order $1/r^{(d/2)-1}$; and for eq. \eqref{Maxwell_AngularMomentumFlux}, to order $1/r^{d/2}$.

{\bf Lorenz gauge and far zone solutions} \qquad To solve eq. \eqref{Maxwell_Eqn} in terms of $A_\mu$ requires a gauge choice, for otherwise the associated wave operator is not invertible. I will choose the Lorentz covariant Lorenz gauge $\partial^\mu A_\mu = 0$, which then translates eq. \eqref{Maxwell_Eqn} into $d$ massless scalar wave equations:
\begin{align}
\label{Maxwell_LorenzGaugeEqns}
\partial^2 A_\nu = J_\nu .
\end{align}
Comparison between eq. \eqref{MasslessScalar_WaveEqn} and \eqref{Maxwell_LorenzGaugeEqns} tells us, the far zone solutions for $A_\nu$ can be simply obtained from equations \eqref{MasslessScalar_EvenDim} and \eqref{MasslessScalar_OddDim} by replacing $\psi \to A_\nu$ and $J \to J_\nu$; whereas those for $F_{\alpha\beta} = \partial_{[\alpha} A_{\beta]}$ may be obtained from equations \eqref{MasslessScalar_PD1_EvenDim} and \eqref{MasslessScalar_PD1_OddDim} by replacing $\psi \to A_\beta$ and $J \to J_\beta$ followed by anti-symmetrizing $\alpha\beta$. 

In even $d=4+2n \geq 4$ dimensions, the far zone vector potential and electromagnetic fields are, respectively,
{\allowdisplaybreaks\begin{align}
		\label{VectorPotential_OddDim}
		A_\nu[t,\vec{x}] 
		&= \frac{1}{2(2\pi r)^{1+n}} \int_{\mathbb{R}^{3+2n}} \dd^{3+2n}\vec{x}' \Bigg( \partial_t^n J_\nu[\tau,\vec{x}'] \\
		&\qquad\qquad
		+ \frac{1}{2} \frac{n(n+1) + (\widehat{r} \cdot \vec{x}') (2n+2) \partial_t - (\vec{x}'^2-(\widehat{r} \cdot \vec{x}')^2) \partial_t^2}{r} \partial_t^{n-1} J_\nu[\tau,\vec{x}'] + \mathcal{O}[r^{-2}] \Bigg) , \nonumber
\end{align}}
with $\tau \equiv t-r+\widehat{r} \cdot \vec{x}'$; and
{\allowdisplaybreaks\begin{align}
		\label{ElectromagneticFields_EvenDim}
		F_{\alpha\beta}[t,\vec{x}] 
		&= \frac{1}{2(2\pi r)^{1+n}} \int_{\mathbb{R}^{3+2n}} \dd^{3+2n}\vec{x}' \Bigg( \left(\delta_{[\alpha}^0 - \delta_{[\alpha}^l \widehat{r}^l\right) \partial_t^{n+1} J_{\beta]}[\tau,\vec{x}'] \\
		&+ \delta_{[\alpha}^a P^{ab} \frac{x'^b}{r} \partial_t^{n+1} J_{\beta]}[\tau,\vec{x}']
		-\frac{n+1}{r} \delta_{[\alpha}^l \widehat{r}^l \partial_t^n J_{\beta]}[\tau,\vec{x}'] \nonumber\\
		&+ \frac{1}{2} \frac{n(n+1) + (\widehat{r} \cdot \vec{x}') (2n+2) \partial_t - (r'^2-(\widehat{r} \cdot \vec{x}')^2) \partial_t^2}{r} \left(\delta_{[\alpha}^0 - \delta_{[\alpha}^l \widehat{r}^l\right) \partial_t^{n} J_{\beta]}[\tau,\vec{x}'] + \mathcal{O}[r^{-2}] \Bigg) . \nonumber
\end{align}}
On the other hand, in odd $d=3+2n \geq 3$ dimensions, the far zone vector potential is
\begin{align}
	\label{VectorPotential_EvenDim}
	A_\nu[t,\vec{x}] 
	&= \frac{1}{\sqrt{2} (2\pi)^{n+1} \cdot r^{n+\frac{1}{2}}} \int_{\mathbb{R}^{2+2n}} \dd^{2+2n}\vec{x}' \int_{0}^{\infty} \dd\mu \exp\left[ - \mu \cdot 0^+ \right]
	\Bigg( \mu^{-\frac{1}{2}} \partial_\tau^n J_\nu\left[ \tau,\vec{x}' \right] \\
	&+ \frac{\mu^{\frac{1}{2}}}{r} \left( \left(n^2 - \frac{1}{4}\right) \partial_\tau^n J_\nu\left[ \tau,\vec{x}' \right] + (\widehat{r} \cdot \vec{x}') (2n+1) \partial_\tau^{n+1} J_\nu\left[ \tau,\vec{x}' \right] - (r'^2-(\widehat{r}\cdot\vec{x}')^2) \partial_\tau^{n+2} J_\nu\left[ \tau,\vec{x}' \right] \right)
	+ \mathcal{O}[r^{-2}] \Bigg) , \nonumber
\end{align}
with $\tau \equiv t - r - \mu + \widehat{r}\cdot\vec{x}'$. And the associated electromagnetic fields are
{\allowdisplaybreaks\begin{align}
		\label{ElectromagneticFields_OddDim}
		F_{\alpha\beta}[t,\vec{x}] 
		&= \frac{1}{\sqrt{2} (2\pi)^{n+1} \cdot r^{n+\frac{1}{2}}} \int_{\mathbb{R}^{2+2n}} \dd^{2+2n}\vec{x}' \int_{0}^{\infty} \dd\mu \exp\left[ - \mu \cdot 0^+ \right] \\
		&\times
		\Bigg\{ \mu^{-\frac{1}{2}} \left( \delta_{[\alpha}^0 - \delta_{[\alpha}^j \widehat{r}^j \right) \partial_\tau^{n+1} J_{\beta]}\left[ \tau,\vec{x}' \right] + \frac{\mu^{-\frac{1}{2}}}{r} \left( \delta_{[\alpha}^{a} P^{ab} x'^b \partial_\tau^{n+1} J_{\beta]}\left[ \tau,\vec{x}' \right]  - \left( n + \frac{1}{2} \right) \widehat{r}^l \delta_{[\alpha}^l \partial_\tau^n J_{\beta]}\left[ \tau,\vec{x}' \right] \right) \nonumber\\
		&\qquad\qquad
		+ \frac{\mu^{-\frac{1}{2}}}{2r} \left( \delta_{[\alpha}^0 - \delta_{[\alpha}^j \widehat{r}^j \right) \Bigg( \left(n^2 - \frac{1}{4}\right) \partial_\tau^{n} J_{\beta]}\left[ \tau,\vec{x}' \right] \nonumber\\
		&\qquad\qquad\qquad\qquad
		+ (\widehat{r} \cdot \vec{x}') (2n+1) \partial_\tau^{n+1} J_{\beta]}\left[ \tau,\vec{x}' \right] - (r'^2-(\widehat{r}\cdot\vec{x}')^2) \partial_\tau^{n+2} J_{\beta]}\left[ \tau,\vec{x}' \right] \Bigg)
		+ \mathcal{O}[r^{-2}] \Bigg\}  . \nonumber
\end{align}}
The integral over $\mu$ in equations \eqref{VectorPotential_OddDim} and \eqref{ElectromagneticFields_OddDim} is the novel feature in odd dimensions relative to that in the even dimensional equations \eqref{VectorPotential_EvenDim} and \eqref{ElectromagneticFields_EvenDim}. In the latter, the signal at $(t,\vec{x})$ was emitted from the electromagnetic current {\it only} at retarded time $t-r+\widehat{r}\cdot\vec{x}'$, due to the Dirac delta function in eq. \eqref{G_EvenDim_FarZone} enforcing strictly-null propagation of the signal. In former, however, the additional integral over $\mu$ describes the fields at $(t,\vec{x})$ as arising from the superposition of waves emitted from the entire past history of the source -- up to retarded time $t-r+\widehat{r}\cdot\vec{x}'$. This history dependence is the signature of inside-the-null cone propagation; and as we saw in \S\eqref{Section_GreenFarZone}, is intimately tied to the presence of inverse fractional powers of frequency in the far zone expansion.

{\bf Non-Relativistic Limit and Current Conservation} \qquad We now assume the electromagnetic current is non-relativistic and proceed to work out in this limit the leading order (i.e., dipolar) contribution to energy-momentum and angular momentum radiated to infinity. To this end, let us first record that current conservation $\partial_\mu J^\mu = 0$ implies the total charge of the system described by $J^\mu$,
\begin{align}
	\label{Maxwell_Charge}
	Q \equiv \int_{\mathbb{R}^{d-1}}\dd^{d-1}\vec{x}' J^0[t,\vec{x}'] ,
\end{align}
is time-independent: $\dot{Q} \equiv \partial_t Q = 0$. (We will see below that electromagnetic energy-momentum radiation begins at the dipole order in all dimensions because the monopole $Q$ is conserved.) Whereas, current conservation implies the dipole
\begin{align}
	\label{Maxwell_SpatialCurrent}
	d^i[t] \equiv \int_{\mathbb{R}^{d-1}} \dd^{d-1}\vec{x}' x'^i J^0[t,\vec{x}']
\end{align}
is related to the total spatial current through its time derivative:
\begin{align}
\dot{d}^i[t] \equiv \partial_\tau d^i[t] 
= \int_{\mathbb{R}^{d-1}} \dd^{d-1}\vec{x}' J^i[t,\vec{x}'] .
\end{align}
We will assume that $J^i \sim J^0 \cdot v$ scales as the charge density $J^0$ multiplied by some characteristic non-relativistic speed $v \ll 1$ describing its internal dynamics; for e.g., $\dot{d} \sim Q \cdot v$. By Taylor expanding the currents in equations \eqref{ElectromagneticFields_EvenDim} and \eqref{ElectromagneticFields_OddDim} in powers of $\widehat{r} \cdot \vec{x}'$, we may associate each factor of $(\widehat{r}\cdot\vec{x}') \partial_t$ in the ensuing expressions to scale as $v$. Higher orders in the Taylor expansion therefore corresponds to a higher order in the non-relativistic expansion.

To leading order in $v$ and for all even $d=4+2n \geq 4$, we may exploit equations \eqref{Maxwell_Charge} and \eqref{Maxwell_SpatialCurrent} to reveal via a direct calculation that the electric fields are
{\allowdisplaybreaks\begin{align}
\label{Electric_EvenDim}
F_{0i}[t,\vec{x}] 
&\approx \frac{1}{2(2\pi r)^{1+n}} \Bigg(  - P^{ij} \partial_t^{n+2} d^{j}[t-r]  \\
&+ \frac{1}{r} \left( \frac{(n+1)(n+2)}{2} \widehat{r}^i \partial_t^n Q
- \frac{n^2 + n + 2}{2} P^{ij} \partial_t^{n+1} d^j[t-r]
+ (2n+2) \widehat{r}^i \widehat{r}^j \partial_t^{n+1} d^j[t-r] \right) \Bigg)  \nonumber		
\end{align}}
while the magnetic fields $F_{ij}$ are
\begin{align}
\label{Magnetic_EvenDim}
F_{ij}[t,\vec{x}]
&\approx \frac{1}{2(2\pi r)^{1+n}} \widehat{r}^{[i} P^{j]a} \left( \partial_t^{n+2} d^{a}[t-r] 
+ \frac{1}{r} \frac{(n+1)(n+2)}{2} \partial_t^{n+1} d^{a}[t-r] \right) . 
\end{align}
A similar calculation would indicate, for all odd $d=3+2n \geq 3$, the leading non-relativistic expansion of the electric fields is
{\allowdisplaybreaks\begin{align}
\label{Electric_OddDim}
F_{0i}[t,\vec{x}] 
&\approx \frac{1}{\sqrt{2} (2\pi)^{n+1} \cdot r^{n+\frac{1}{2}}} \int_{0}^{\infty} \dd\mu \exp\left[ - \mu \cdot 0^+ \right] \\
&\times \Bigg\{ - \mu^{-\frac{1}{2}} P^{ij} \partial_t^{n+2} d^{j}\left[ t-r-\mu \right]
+ \frac{\mu^{-\frac{1}{2}}}{r} \Bigg( \left( \frac{n^2}{2} + n + \frac{3}{8} \right) \widehat{r}^i \partial_t^n Q + (2n+1) \widehat{r}^i \widehat{r}^j \partial_t^{n+1} d^j\left[t-r-\mu\right]  \nonumber\\
&\qquad\qquad
- \left(\frac{n^2}{2} + \frac{7}{8} \right) P^{ij} \partial_t^{n+1} d^j\left[t-r-\mu\right] \Bigg) \Bigg\} . \nonumber
\end{align}}
and that of the magnetic fields is
{\allowdisplaybreaks\begin{align}
\label{Magnetic_OddDim}
F_{ij}[t,\vec{x}]
&\approx \frac{1}{\sqrt{2} (2\pi)^{n+1} \cdot r^{n+\frac{1}{2}}} \int_{0}^{\infty} \dd\mu \exp\left[ - \mu \cdot 0^+ \right] \\
&\times 
\Bigg\{ \mu^{-\frac{1}{2}} \widehat{r}^{[i} \partial_t^{n+2} d^{j]}\left[ t-r-\mu \right] + \frac{\mu^{-\frac{1}{2}}}{r} \left( \frac{n^2}{2} + n + \frac{3}{8} \right) \widehat{r}^{[i} \partial_t^{n+1} d^{j]}\left[ t-r-\mu \right] \Bigg\} . \nonumber
\end{align}}
I have checked that equations \eqref{Electric_EvenDim} through \eqref{Magnetic_OddDim} satisfy Maxwell's equations $\partial_i F_{0i} = 0$ and $\partial_0 F_{0j} = \partial_i F_{ij}$. I also highlight here, it is important {\it not} to take the non-relativistic limit -- replacing $t-r+\widehat{r}\cdot\vec{x}'$ with $t-r$ -- too early. For instance, in 4D, if we take from the outset the vector potential as its leading order expression, $A_\nu \approx (4\pi r)^{-1} \int_{\mathbb{R}^3} \dd^{3}\vec{x}' J_{\nu}[t-r,\vec{x}']$, a quick calculation would reveal the Lorenz gauge is violated $(\partial^\nu A_\nu \neq 0)$ and therefore current conservation cannot be exploited consistently to re-write this $A_\nu$ and its associated $F_{\alpha\beta}$ into dipole moments.

To reiterate: equations \eqref{Electric_EvenDim} through \eqref{Magnetic_OddDim} have been expanded up to the dipolar order, with relative corrections that scale as $v \sim (\widehat{r} \cdot \vec{x}') \partial_t$; as well as both the $1/r^{(d/2)-1}$ and $1/r^{d/2}$ orders in the far zone expansion.

{\bf Dipolar Energy-Momentum Radiation} \qquad Inserting the leading order $1/r^{(d/2)-1}$ even dimensional electromagnetic fields of equations \eqref{Electric_EvenDim} and \eqref{Magnetic_EvenDim} and odd dimensional ones of equations \eqref{Electric_OddDim} and \eqref{Magnetic_OddDim} into eq. \eqref{Maxwell_EnergyFlux} now hands us the corresponding dipole radiation formulas. 

For all even $d \geq 4$,
{\allowdisplaybreaks\begin{align}
T^{0i}	
&= \frac{\widehat{r}^i}{4(2\pi r)^{d-2}} \left( P^{ab} \partial_t^{\frac{d}{2}} d^b[t-r] \right)^2 \left(  1 + \mathcal{O}[r^{-1}] \right) , 
\end{align}
Because $P_{ia} P_{ib} = P_{ab}$,
\begin{align}
\label{Maxwell_DipoleEnergyAngularRadiation_EvenDim}
\frac{\dd E}{\dd t \dd \Omega}
&= \frac{1}{4(2\pi)^{d-2}} \left( P^{ab} \partial_t^{\frac{d}{2}} d^b[t-r] \right)^2 \\
&= \frac{\sin^2\vartheta}{4(2\pi)^{d-2}} \left( \partial_t^{\frac{d}{2}} \vec{d}[t-r] \right)^2 ; \nonumber
\end{align}}
where $P^{ab} \partial_t^{\frac{d}{2}} d^b[t-r] = (\delta^{ab} - \widehat{r}^a \widehat{r}^b) \partial_t^{\frac{d}{2}} d^b[t-r]$ is simply the transverse part of the dipole at retarded time $t-r$ and the $\vartheta$ is the angle between the radial direction (to the observer) and the $(d/2)$th time derivative of the retarded dipole:
\begin{align}
\left\vert \partial_t^{\frac{d}{2}} \vec{d}[t-r] \right\vert \cdot \cos\vartheta 
\equiv \widehat{r} \cdot \partial_t^{\frac{d}{2}} \vec{d}[t-r] .
\end{align}
For all odd $d \geq 3$, on the other hand,
\begin{align}
	T^{0i}
	&= \frac{\widehat{r}^i}{2(2\pi)^{d-1} r^{d-2}} \left( P^{ab} \int_{0}^{\infty} \frac{\dd\mu}{\mu^{\frac{1}{2}}} \partial_t^{\frac{d+1}{2}} d^b[t-r-\mu] \right)^2  \left( 1 + \mathcal{O}[r^{-1}] \right) . \\
	\label{Maxwell_DipoleEnergyAngularRadiation_OddDim}
	\frac{\dd E}{\dd t \dd \Omega}
	&= \frac{1}{2(2\pi)^{d-1}} \left( P^{ab} \int_{0}^{\infty} \frac{\dd\mu}{\mu^{\frac{1}{2}}} \partial_t^{\frac{d+1}{2}} d^b[t-r-\mu] \right)^2 \\
	&= \frac{\sin^2\vartheta}{2(2\pi)^{d-1}} \left( \int_{0}^{\infty} \frac{\dd\mu}{\mu^{\frac{1}{2}}} \partial_t^{\frac{d+1}{2}} \vec{d}[t-r-\mu] \right)^2 , \nonumber
\end{align}
with $\vartheta$ defined as the angle between the radial direction (to the observer) and the $(1/2)(d+1)$th time derivative of the retarded history integral of the dipole:
\begin{align}
	\left\vert \int_{0}^{\infty} \frac{\dd\mu}{\mu^{\frac{1}{2}}} \partial_t^{\frac{d+1}{2}} \vec{d}[t-r-\mu] \right\vert \cdot \cos\vartheta 
	\equiv \widehat{r} \cdot \int_{0}^{\infty} \frac{\dd\mu}{\mu^{\frac{1}{2}}} \partial_t^{\frac{d+1}{2}} \vec{d}[t-r-\mu] .
\end{align}
By appealing to the solid angle tensor integrals in equations \eqref{SolidAngle} and \eqref{TensorIntegral_2PointFunction} below, we may integrate both sides of equations \eqref{Maxwell_DipoleEnergyAngularRadiation_EvenDim} and \eqref{Maxwell_DipoleEnergyAngularRadiation_OddDim} over the $(d-2)-$sphere and arrive at the dipole radiation formula in equations \eqref{Maxwell_DipoleEnergyRadiation_EvenDim} and \eqref{Maxwell_DipoleEnergyRadiation_OddDim}.

Here and below, we derive the electromagnetic and gravitational energy and angular momentum lost to infinity per unit time; i.e., $\dd E/\dd t$ and $\dd L^{ij}/\dd t$. Because they are quadratic in the dipole or quadrupole moments, we may integrate these expressions over all time and invoke Parseval's theorem to re-express the total energy or angular momentum loss as an integral over all angular frequencies. For instance, starting from eq. \eqref{Maxwell_DipoleEnergyRadiation_EvenDim}:
\begin{align}
\int_{\mathbb{R}} \dd t \frac{\dd E}{\dd t}  
&= \frac{d-2}{2^{d} \pi^{\frac{d-3}{2}} \Gamma[\frac{d+1}{2}]} \int_{\mathbb{R}} \frac{\dd\omega}{2\pi} (-i \omega)^{\frac{d}{2}} \widetilde{d}^a[\omega] \cdot (+i \omega)^{\frac{d}{2}} \widetilde{d}^a[-\omega] ,
\end{align}
where $\widetilde{d}^a[\omega]$ denotes the Fourier coefficient of the real-time dipole moment $d^a[t]$. This allows us to interpret the integrand of the right hand side as the rate of energy loss per unit angular frequency, and thereby arrive at eq. \eqref{Maxwell_DipoleEnergyRadiation_FreqSpace}. For the odd dimensional case in eq. \eqref{Maxwell_DipoleEnergyRadiation_OddDim}, a similar argument applies, if we first recognize the frequency decomposition
\begin{align}
	\int_{0}^{\infty} \frac{\dd \mu}{\sqrt{\mu}} e^{-\mu \cdot 0^+} d^j[t-r-\mu]
	&= \int \frac{\dd\omega}{2\pi} \frac{\sqrt{\pi}}{(-i (\omega + i 0^+))^{\frac{1}{2}}} \exp\left[-i \omega (t-r)\right] \widetilde{d}^j[\omega] .
\end{align}
Even though the real time electromagnetic dipole radiation formula takes a different form in even versus odd dimensions; we see its frequency space formulas take the same form for all dimensions.

I further highlight here: by referring to the leading order terms in equations \eqref{Electric_EvenDim} through \eqref{Magnetic_OddDim}, the factors within the parenthesis in equations \eqref{Maxwell_DipoleEnergyAngularRadiation_EvenDim} and \eqref{Maxwell_DipoleEnergyAngularRadiation_OddDim} are readily seen to be the time-dependent radiative pieces of the electromagnetic fields. These will play a central role below in determining the signal of our `light bulb' as seen by a distant observer.

{\bf Dipolar Angular Momentum Radiation} \qquad We now turn to computing the net angular momentum radiated to infinity, by employing eq. \eqref{Maxwell_StressTensor} in eq. \eqref{AngularMomentumFlux}:
\begin{align}
\frac{\dd L^{ij}}{\dd t \dd\Omega}
	&= \lim_{r \to \infty} r^{d-2} \left( - x^{[i} F_{j]0} F_{k0} \widehat{r}^k + x^{[i} F^{j]l} F^{kl} \widehat{r}^k \right) .
\end{align}
The extra $r$ arising from the $\vec{x}$ factor in this angular momentum calculation, relative to the energy-momentum flux in eq. \eqref{EnergyFlux}, is why we need to develop $F_{\alpha\beta}$ one order in $1/r$ beyond the leading $1/r^{(d/2)-1}$. Note that, a direct calculation would show that the rightmost term of eq. \eqref{Maxwell_AngularMomentumFlux_Step1} goes as $\widehat{r}^{[i} F^{j]l} F^{kl} \widehat{r}^k \propto \widehat{r}^l F_{kl} \widehat{r}^k$, and thus drops out of the final result due to the anti-symmetry of $F_{kl}$, so we really have
\begin{align}
	\label{Maxwell_AngularMomentumFlux_Step1}
	\frac{\dd L^{ij}}{\dd t \dd\Omega}
	&= - \lim_{r \to \infty} r^{d-2} x^{[i} F_{j]0} F_{k0} \widehat{r}^k .
\end{align}
In even dimensions $d=4+2n \geq 4$, the rate of flow of the $(i,j)$ generator of rotations to infinity is
{\allowdisplaybreaks\begin{align}
\label{Maxwell_AngularMomentumFlux_EvenDim_PerOmega}
\frac{\dd L^{ij}}{\dd t \dd \Omega}
&= \frac{n+1}{2(2\pi)^{2+2n}} \widehat{r}^{[i} \partial_t^{n+2} d^{j]}[t-r] \widehat{r}^{a} \partial_t^{n+1} d^{a}[t-r] ;
\end{align}}
where I have discarded the term
\begin{align}
	\label{Maxwell_AngularMomentumFlux_EvenDim_PerOmega_Dropped}
	\frac{1}{4(2\pi)^{2+2n}} \widehat{r}^{[i} \partial_t^{n+2} d^{j]}[t-r] \frac{(n+1)(n+2)}{2} \partial_t^n Q 
\end{align}
by assuming that
\begin{align}
\left. \widehat{r}^{[i} \partial_t^{n+1} d^{j]}[t-r] \right\vert_{t = -\infty}^{t = \infty} = 0 ;
\end{align}
so that eq. \eqref{Maxwell_AngularMomentumFlux_EvenDim_PerOmega_Dropped} does not contribute to the {\it total} loss of angular momentum. In any case, eq. \eqref{Maxwell_AngularMomentumFlux_EvenDim_PerOmega_Dropped} is non-zero only in $d=4$. 

Exploiting equations \eqref{SolidAngle} and \eqref{TensorIntegral_2PointFunction} below to integrate eq. \eqref{Maxwell_AngularMomentumFlux_EvenDim_PerOmega} over solid angle,
{\allowdisplaybreaks\begin{align}
\label{Maxwell_AngularMomentumFlux_EvenDim_Total}
\frac{\dd L^{ij}}{\dd t}
&= \frac{d-2}{2^{d} \pi^{\frac{d-3}{2}} \Gamma[\frac{d+1}{2}]} \partial_t^{\frac{d-2}{2}} d^{[i}[t-r] \partial_t^{\frac{d}{2}} d^{j]}[t-r] .
\end{align}}
In 4D, this formula recovers the spatial Hodge dual of the result in Jackson Problem 9.9 \cite{Jackson}. 
{\allowdisplaybreaks\begin{align}
\left. \frac{\dd L^{ij}}{\dd t} \right\vert_{d=4}
		&= \frac{1}{6 \pi} \partial_t d^{[i}[t-r] \partial_t^{2} d^{j]}[t-r]
\end{align}}
In odd dimensions $d=3+2n \geq 3$, on the other hand,
{\allowdisplaybreaks\begin{align}
\label{Maxwell_AngularMomentumFlux_OddDim_PerOmega}
\frac{\dd L^{ij}}{\dd t \dd \Omega}
		&= \frac{2n+1}{2 (2\pi)^{2n+2}} \int_{0}^{\infty} \frac{\dd\mu}{\sqrt{\mu}} e^{- \mu \cdot 0^+} \widehat{r}^{[i} \partial_t^{n+2} d^{j]}\left[ t-r-\mu \right] 
		\int_{0}^{\infty} \frac{\dd\mu'}{\sqrt{\mu'}} e^{- \mu' \cdot 0^+} \widehat{r}^a \partial_t^{n+1} d^a\left[t-r-\mu'\right] ;
\end{align}}
where I have dropped the (potentially divergent) total time derivative term
{\allowdisplaybreaks\begin{align}
		\label{Maxwell_AngularMomentumFlux_OddDim_PerOmega_Dropped}
		&\frac{1}{2 (2\pi)^{2n+2}} \int_{0}^{\infty} \frac{\dd\mu}{\sqrt{\mu}} e^{- \mu \cdot 0^+} 
		\widehat{r}^{[i} \partial_t^{n+2} d^{j]}\left[ t-r-\mu \right] \int_{0}^{\infty} \frac{\dd\mu'}{\sqrt{\mu'}} e^{- \mu' \cdot 0^+}
		\left( \frac{n^2}{2} + n + \frac{3}{8} \right) \partial_t^n Q .
\end{align}}
This amounts to assuming both the lower limit term
\begin{align}
\left. \widehat{r}^{[i} \partial_t^{n+1} d^{j]}\left[ t-r-\mu \right] \right\vert_{t=-\infty}
\end{align}
and upper limt integral
\begin{align}
\lim_{t \to \infty} \int_{-t}^{\infty} \frac{\dd\mu'}{\sqrt{\mu'+t}} e^{- (\mu'+t) \cdot 0^+} 
\widehat{r}^{[i} \partial_t^{n+1} d^{j]}\left[ -r-\mu' \right]
\end{align}
can both be set to zero. Of course, a more careful analysis is needed to understand the (physical) origin of the divergent integral in eq. \eqref{Maxwell_AngularMomentumFlux_OddDim_PerOmega_Dropped}, as well as how to handle it.

Integrating eq. \eqref{Maxwell_AngularMomentumFlux_OddDim_PerOmega} over the solid angle,
{\allowdisplaybreaks\begin{align}
		\label{Maxwell_AngularMomentumFlux_OddDim_Total}
		\frac{\dd L^{ij}}{\dd t}
		&= \frac{d-2}{2^{d} \pi^{\frac{d-1}{2}} \Gamma[\frac{d+1}{2}]} \int_{0}^{\infty} \frac{\dd\mu'}{\sqrt{\mu'}} e^{-\mu' \cdot 0^+} \partial_t^{\frac{d-1}{2}} d^{[i}\left[t-r-\mu'\right] \cdot
		\int_{0}^{\infty} \frac{\dd\mu}{\sqrt{\mu}} e^{-\mu \cdot 0^+} \partial_t^{\frac{d+1}{2}} d^{j]}\left[ t-r-\mu \right] .
\end{align}}
Transforming the real time equations \eqref{Maxwell_AngularMomentumFlux_EvenDim_Total} and \eqref{Maxwell_AngularMomentumFlux_OddDim_Total} into frequency space, we discover they -- like their energy flux counterparts -- now take the same form for all $d \geq 3$.
{\allowdisplaybreaks\begin{align}
\frac{\dd L^{ij}}{\dd \omega}
		&= \frac{d-2}{2^{d+1} \pi^{\frac{d-1}{2}} \Gamma[\frac{d+1}{2}]} \omega^{d-1}
		\left( \widetilde{d}^{[i}\left[\omega\right] \cdot \widetilde{d}^{j]}\left[ \omega \right]^* \right) 
\end{align}}
{\bf `Light Bulb' In Even Versus Odd Dimensions} \qquad In the final portion of this section on electromagnetic radiation, let us attempt to answer the question posed in the abstract: 
\begin{quotation}
``How long does a light bulb shine in odd dimensional flat spacetime, according to a distant observer?"
\end{quotation}
I will suppose that a light bulb may be modeled as a incoherent collection of dipoles oscillating over a range of frequencies. As a first pass to this problem, therefore, I will focus on the propagation of the electromagnetic signal generated by a single dipole at a fixed frequency. Furthermore, if the dipole were active {\it only} over a finite duration $0 \leq t \leq T$, the key time-dependent factors in equations \eqref{Maxwell_DipoleEnergyAngularRadiation_EvenDim} and \eqref{Maxwell_DipoleEnergyAngularRadiation_OddDim} may be described as
\begin{align}
\label{LightBulb_DipoleEvenDim}
\partial_t^{\frac{d}{2}} \vec{d}[0 \leq t \leq T] 
	&= \vec{C}_\ell \cdot \frac{\sin\left[ \omega_\ell t \right]}{\sqrt{T/2}} ,
	\qquad\qquad \text{(Even $d \geq 4$)} , \\
\label{LightBulb_DipoleOddDim}
\partial_t^{\frac{d+1}{2}} \vec{d}[0 \leq t \leq T] 
	&= \vec{C}_\ell \cdot \frac{\sin\left[ \omega_\ell t \right]}{\sqrt{T/2}} ,
	\qquad\qquad \text{(Odd $d \geq 3$)} ;
\end{align}
where $\omega_\ell \equiv \pi\ell/T$, and zero for $t$ outside this interval. These $\{ \sqrt{2/T} \sin[\omega_\ell t] | \ell=1,2,3,\dots \}$ are orthonormal basis functions that vanishes at $t=0$ and $t=T$, and the $\{ C_\ell^i \}$ are arbitrary coefficients. It is certainly possible to imagine that a collection of incoherent dipoles -- when averaged $\langle \dots \rangle$ over their coherence times -- would produce
\begin{align}
\langle \partial_t^{\frac{d}{2}} d^i[0 \leq t \leq T] \partial_t^{\frac{d}{2}} d^j[0 \leq t' \leq T] \rangle 
	&= \sum_{\ell=1}^\infty C_\ell^i C_\ell^j \cdot \frac{2}{T} \sin\left[ \omega_\ell t \right] \sin\left[ \omega_\ell t' \right] ,
	\qquad\qquad \text{(Even $d \geq 4$)} , \\
\langle \partial_t^{\frac{d+1}{2}} d^i[0 \leq t \leq T] \partial_t^{\frac{d+1}{2}} d^j[0 \leq t' \leq T] \rangle
	&= \sum_{\ell=1}^\infty C_\ell^i C_\ell^j \cdot \frac{2}{T} \sin\left[ \omega_\ell t \right] \sin\left[ \omega_\ell t' \right] ,
	\qquad\qquad \text{(Odd $d \geq 3$)} ;
\end{align}
so that the total signal would be the superposition from each distinct frequency, without any cross terms.

With this in mind, let us return to the single-frequency emitter of equations \eqref{LightBulb_DipoleEvenDim} and \eqref{LightBulb_DipoleOddDim}. We see that the electromagnetic energy flux at a large distance in even $d \geq 4$, according to eq. \eqref{Maxwell_DipoleEnergyAngularRadiation_EvenDim}, is
{\allowdisplaybreaks\begin{align}
\frac{\dd E}{\dd t \dd \Omega}
&= \frac{C_\ell^a P^{ab} C_\ell^b}{2(2\pi)^{d-2} T} \mathcal{E}[\omega_\ell(t-r)]^2, & \\
\label{LightBulb_EvenDim_Active}
\mathcal{E}[\omega_\ell(t-r)]
&= \sin \left[\omega_\ell(t-r)\right] , \qquad &0 \leq t-r \leq T  \\
&= 0 \qquad &\text{otherwise} .
\end{align}}
Just like the 4D case, if the `light bulb' is turned on for a duration $T$ in even dimensions, the distant observer will see it for the exactly the same duration $T$; the main differences from the 4D case are the $1/$(observer-source distance)$^{d-2}$ dimension-dependent power law fall-off as well as the number of time derivatives acting on the dipole driving the radiation.

The odd dimensional case is slightly more involved. Inserting eq. \eqref{LightBulb_DipoleOddDim} into the history integral of eq. \eqref{Maxwell_DipoleEnergyAngularRadiation_OddDim},
\begin{align}
\int_{0}^{\infty} \frac{\dd\mu}{\mu^{\frac{1}{2}}} \partial_t^{\frac{d+1}{2}} d^i[t-r-\mu]
	&= 0 				& t-r<0 \\
	&= C^i_\ell \cdot \int_{0}^{t-r} \frac{\dd\mu}{\sqrt{\mu}} \frac{\sin\left[ \omega_\ell (t-r-\mu) \right]}{\sqrt{T/2}} & 0<t-r<T \\
	&= C^i_\ell \cdot \int_{t-r-T}^{t-r} \frac{\dd\mu}{\sqrt{\mu}} \frac{\sin\left[ \omega_\ell (t-r-\mu) \right]}{\sqrt{T/2}} & t-r>T .
\end{align}
These integrals are related to the Fresnel cosine and sine integrals, which I shall dub respectively as FC and FS. If we write 
\begin{align}
\label{LightBulb_OddDim_Active}
\frac{\dd E}{\dd t \dd \Omega}
&= \frac{C_\ell^a P^{ab} C_\ell^b}{(2\pi)^{d-2} (\omega_\ell T)} \mathcal{E}[\omega_\ell(t-r)]^2 , 
\end{align}
Then, for retarded time $t-r$ lying within the active duration,
\begin{align}
\label{LightBulb_OddDim_Active_E}
\mathcal{E}[0 < \omega_\ell(t-r) < \omega_\ell T]
&\equiv \sin\left[ \omega_\ell (t-r) \right] \text{FC}\left[ \sqrt{\frac{2\omega_\ell(t-r)}{\pi}} \right] - \cos\left[ \omega_\ell (t-r) \right] \text{FS}\left[ \sqrt{\frac{2\omega_\ell(t-r)}{\pi}} \right] ;
\end{align}
whereas for $t-r$ after the active duration,
\begin{align}
\label{LightBulb_OddDim_AfterActive_E}
\mathcal{E}[\omega_\ell(t-r) > \omega_\ell T]
&\equiv
\sin\left[ \omega_\ell (t-r) \right] \left\{ \text{FC}\left[ \sqrt{\frac{2\omega_\ell(t-r)}{\pi}} \right] - \text{FC}\left[ \sqrt{\frac{2\omega_\ell(t-r-T)}{\pi}} \right] \right\} \\
&\qquad\qquad\qquad\qquad
- \cos\left[ \omega_\ell (t-r) \right] \left\{ \text{FS}\left[ \sqrt{\frac{2\omega_\ell(t-r)}{\pi}} \right] - \text{FS}\left[ \sqrt{\frac{2\omega_\ell(t-r-T)}{\pi}} \right] \right\} . \nonumber
\end{align}
\begin{figure}[!ht]
	\begin{center}
		\includegraphics[width=5in]{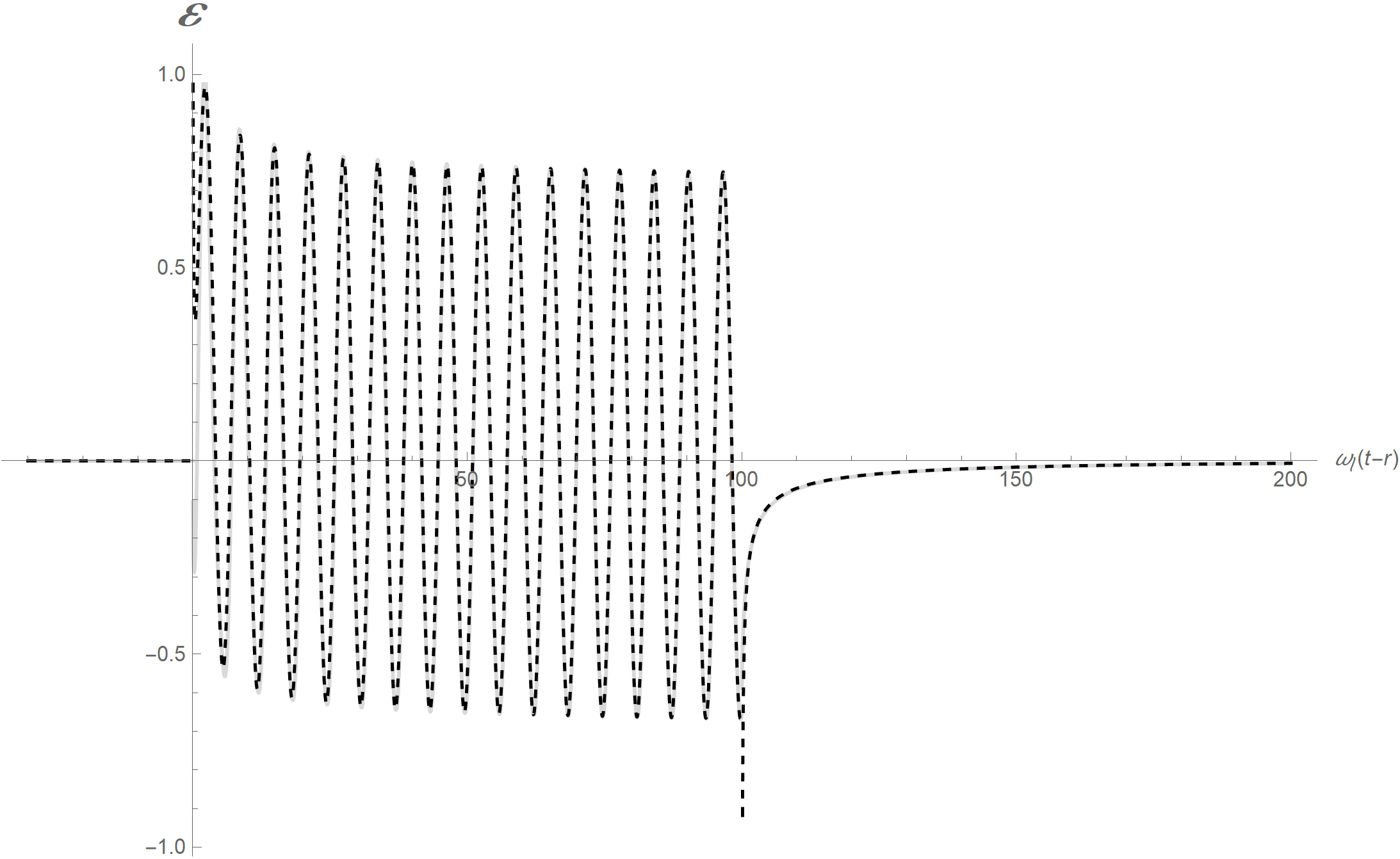} \\ 
		\includegraphics[width=2.5in]{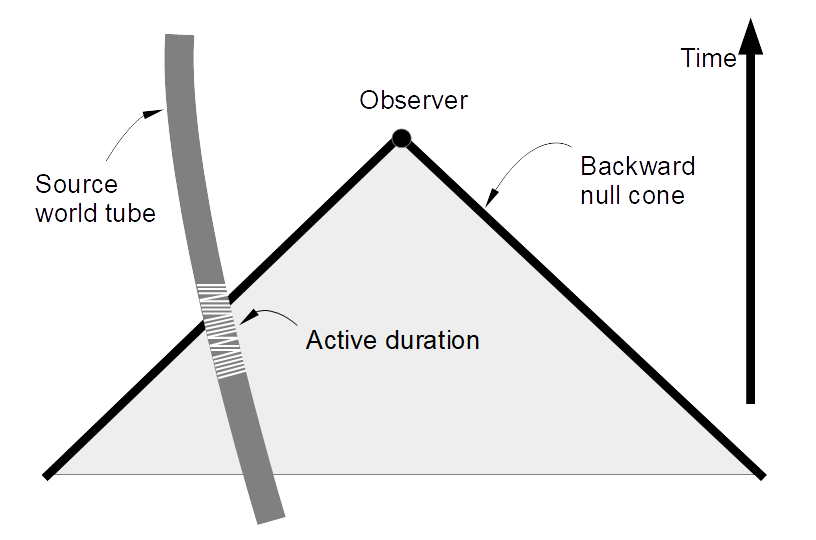} 
		\includegraphics[width=2.5in]{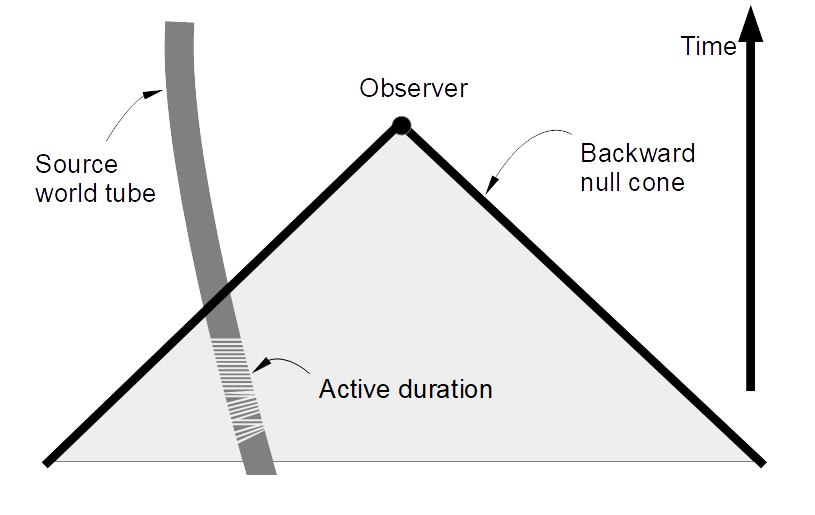} 
		\caption{{\it Top panel: }The time-dependent part ($\mathcal{E}[\omega_\ell(t-r)]$) of the leading order electromagnetic radiation field, for a `light bulb' of duration $\omega_\ell T = 100$, as a function of retarded time $\omega_\ell(t-r)$. The bold line is the exact time dependence in equations \eqref{LightBulb_OddDim_Active_E} and \eqref{LightBulb_OddDim_AfterActive_E}; whereas the dashed line is their asymptotic expansion in equations \eqref{LightBulb_OddDim_Active_Asymptotic} and \eqref{LightBulb_OddDim_AfterActive_Asymptotic}. We see that the gradual downward shift of the sinusoidal waveform is due to the $1/\sqrt{\omega_\ell(t-r)}$ in eq. \eqref{LightBulb_OddDim_Active_Asymptotic}; whereas according to eq. \eqref{LightBulb_OddDim_AfterActive_Asymptotic}, after the active duration the signal decays to zero rather quickly and without any oscillations. Note that the energy flux $\dd E/(\dd t \dd \Omega)$ of the `light bulb' in eq. \eqref{LightBulb_DipoleOddDim} is proportional to the {\it square} of the bold line of this plot (cf. eq. \eqref{LightBulb_OddDim_Active}). {\it Bottom left panel: }Spacetime configuration of the observer receiving signals from the source when her retarded time $t-r$ lies within its active duration (dashed segment). This corresponds to the $0 \leq \omega_\ell(t-r) \leq 100$ region of the plot in the top panel. {\it Bottom right panel: }Spacetime configuration of the observer receiving signals from the source after her retarded time $t-r$ has passed its active duration (dashed segment). This corresponds to the $\omega_\ell(t-r) > 100$ region of the plot in the top panel.}
		\label{LightBulbInOddDim}
	\end{center}
\end{figure}

To gain some insight into the behavior of this signal, let us first record the large argument asymptotic expansions of FS and FC:
{\allowdisplaybreaks\begin{align}
		\text{FC}[z \gg 1] 
		&\sim \frac{1}{2} + \frac{\sin\left[ \frac{\pi z^2}{2} \right]}{\pi z} \sum_{m=0}^{\infty} \frac{(-)^m \left(1/2\right)_{2m}}{ (\pi z^2/2)^{2m} }
		- \frac{\cos\left[ \frac{\pi z^2}{2} \right]}{\pi z} \sum_{m=0}^{\infty} \frac{(-)^m \left(1/2\right)_{2m+1}}{ (\pi z^2/2)^{2m+1} } , \\
		\text{FS}[z \gg 1] 
		&\sim \frac{1}{2} - \frac{\cos\left[ \frac{\pi z^2}{2} \right]}{\pi z} \sum_{m=0}^{\infty} \frac{(-)^m \left(1/2\right)_{2m}}{ (\pi z^2/2)^{2m} }
		- \frac{\sin\left[ \frac{\pi z^2}{2} \right]}{\pi z} \sum_{m=0}^{\infty} \frac{(-)^m \left(1/2\right)_{2m+1}}{ (\pi z^2/2)^{2m+1} } ,
\end{align}}
where $(a)_{n}$ is the Pochhammer symbol. Utilizing them in equations \eqref{LightBulb_OddDim_Active_E} and \eqref{LightBulb_OddDim_AfterActive_E} teaches us, whenever $\omega_\ell (t-r) \gg 1$, then to leader order in $1/\sqrt{\omega_\ell(t-r)}$, the time dependent portion of the electromagnetic field goes as
\begin{align}
\label{LightBulb_OddDim_Active_Asymptotic}
\mathcal{E}[0 < \omega_\ell(t-r) < \omega_\ell T] 
&\sim \frac{\sin\left[ \omega_\ell (t-r) - \frac{\pi}{4} \right]}{\sqrt{2}} 
+ \frac{1}{\sqrt{2 \pi \omega_\ell (t-r)}} 
\end{align}
and
\begin{align}
	\label{LightBulb_OddDim_AfterActive_Asymptotic}
	\mathcal{E}[\omega_\ell(t-r) > \omega_\ell T]
	&\sim
	\frac{1}{\sqrt{2 \pi \omega_\ell (t-r)}} 
	- \frac{\cos\left[ \omega_\ell T \right]}{\sqrt{2 \pi \omega_\ell (t-r-T)}}  . 
\end{align} 
In Fig. \eqref{LightBulbInOddDim}, I compare the exact expressions in equations \eqref{LightBulb_OddDim_Active_E} and \eqref{LightBulb_OddDim_AfterActive_E} (bold line) to their asymptotic forms in equations \eqref{LightBulb_OddDim_Active_Asymptotic} and \eqref{LightBulb_OddDim_AfterActive_Asymptotic} (dashed line): except for the narrow regions around the transitions $\omega_\ell (t-r) \sim 0$ and $\omega_\ell(t-r) \sim \omega_\ell T$, we see the latter describes the electromagnetic field very well.

As alluded to earlier, the asymptotic results of integration $\mathcal{E}$ occurring in equations \eqref{LightBulb_OddDim_Active_Asymptotic} and \eqref{LightBulb_OddDim_AfterActive_Asymptotic} arise from the time-dependent factors of the electric and magnetic fields. When the retarded time lies within the time period when the light bulb is lit, the observer will see an electromagnetic field with the same frequency as that of the $(1/2)(d+1)$th derivative of the dipole, except there is a phase shift of $-\pi/4$. This sinusoidal signal suffers a `DC' shift that decays at leading order with retarded time as $1/\sqrt{\omega_\ell(t-r)}$. By contrasting these results against the even dimensional case in eq. \eqref{LightBulb_EvenDim_Active}, we see the phase shift and additional power law decay are the tail induced features unique to odd dimensions.

After the light bulb has been turned off, $t-r>T$, the remaining electromagnetic field strength -- which is then pure tail -- is no longer oscillatory but is the superposition of several power law decays. This qualitative behavior indicates, even though the pure tail signal after the bulb has ceased is non-trivial, it does decay away fairly rapidly. This in turn verifies the assertion that the electromagnetic signal of a light bulb of finite duration $T$ as detected by a distant observer is roughly of the same duration $T$, because the tail signal in eq. \eqref{LightBulb_OddDim_AfterActive_Asymptotic} quickly tends to zero.

\section{Weak Field $\Lambda=0$ Limits of Einstein's General Relativity}
\label{Section_Gravity}

{\bf Setup} \qquad I now move on to perform a similar analysis for the weak field limit of Einstein's General Relativity without a cosmological constant ($\Lambda=0$); i.e.,
\begin{align}
	\label{GR}
	G_{\mu\nu} = 8 \pi \GN T_{\mu\nu} ,
\end{align}
off a flat spacetime background
\begin{align}
	g_{\mu\nu} = \eta_{\mu\nu} + h_{\mu\nu} , \qquad |h_{\mu\nu}| \ll 1.
\end{align}
For concrete calculations, it is convenient to perform a change of field variables to
\begin{align}
\bar{h}_{\mu\nu} = h_{\mu\nu} - \frac{1}{2} \eta_{\mu\nu} h ,
\end{align}
with the reverse transformation being
\begin{align}
\label{hbToh}
h_{\mu\nu} = \bar{h}_{\mu\nu} - \frac{2}{d-2} \eta_{\mu\nu} \bar{h} .
\end{align}
Throughout the rest of this paper, whenever we are carrying out gravitational perturbation theory, the indices on the perturbations (either $\bar{h}_{\mu\nu}$ or $h_{\mu\nu}$) and on the partial derivative $\partial_\mu$ will be moved with the flat metric. For example, the inverse metric is a geometric series:
\begin{align}
g^{\mu\nu} = \eta^{\mu\nu} - h^{\mu\sigma} h_\sigma^{\phantom{\sigma}\nu} + h^{\mu\sigma} h_\sigma^{\phantom{\sigma}\rho} h_\rho^{\phantom{\rho}\nu} + \mathcal{O}[h^4] . 
\end{align}
If we denote $\bar{T}_{\mu\nu}$ as the portion of the energy-momentum-shear-stress tensor of matter without any $\bar{h}_{\mu\nu}$; and $\delta_n G_{\mu\nu}$ and $\delta_n T_{\mu\nu}$ respectively as the piece of Einstein's tensor $G_{\mu\nu}$ and the energy-stress tensor of matter $T_{\mu\nu}$ containing exactly $n \geq 1$ powers of the metric perturbation $\bar{h}_{\mu\nu}$, then General Relativity in eq. \eqref{GR} itself can be re-expressed as an infinite series:
\begin{align}
\label{GR_SeriesForm}
\delta_1 G_{\mu\nu} 
&= 8\pi\GN \left( \bar{T}_{\mu\nu} + \sum_{n=1}^{+\infty} \delta_n T_{\mu\nu}
+ \sum_{n=2}^{+\infty} \delta_n t_{\mu\nu} \right) , \\
\label{GR_GWtmunu}
\delta_n t_{\mu\nu}
&\equiv -\frac{\delta_n G_{\mu\nu}}{8\pi\GN} .
\end{align}
The form of General Relativity in eq. \eqref{GR_SeriesForm} admits the following interpretation. Away from the matter source, and particularly in the $r \to \infty$ far zone where $T_{\mu\nu}=0$, we may associate the right hand side with the energy-momentum tensor of gravity itself in a Minkowski spacetime -- for, as I will now argue, it is not only symmetric, it is also divergence-free with respect to the flat metric when $\bar{h}_{\mu\nu}$ satisfies its equations-of-motion. Firstly, just like Maxwell's electromagnetism, to solve Einstein's equations (perturbatively) one has to fix a gauge so that the associated wave operator is invertible. We shall use the de Donder gauge
\begin{align}
	\label{deDonder}
	\partial^\mu \bar{h}_{\mu\nu} = 0 ,
\end{align}
so that the linearized Einstein's tensor becomes $\delta_1 G_{\mu\nu} = -(1/2) \partial^2 \bar{h}_{\mu\nu}$ and eq. \eqref{GR_SeriesForm} now reads
\begin{align}
	\label{GR_SeriesForm_deDonder}
	\partial^2 \bar{h}_{\mu\nu} 
	&= -16\pi\GN \left( \bar{T}_{\mu\nu} + \sum_{n=1}^{+\infty} \delta_n T_{\mu\nu}
	+ \sum_{n=2}^{+\infty} \delta_n t_{\mu\nu} \right) .
\end{align}
Now, the left hand side is no longer identically divergence-free even though the non gauge fixed $\delta_1 G_{\mu\nu}$ is; but rather, its divergence -- and thus that of the right hand side -- is zero when the $\bar{h}_{\mu\nu}$ satisfies its equations-of-motion (aka ``on-shell") and hence the de Donder gauge. In the $r \to \infty$ far zone, we conclude
\begin{align}
\partial^\mu \sum_{n=2}^{+\infty} \delta_n t_{\mu\nu} = 0 \qquad \text{(whenever $\bar{h}_{\alpha\beta}$ is on shell)} .
\end{align}
The $\bar{h}_{\mu\nu}$ itself may be solved by multiplying $1/\partial^2$ on both sides of eq. \eqref{GR_SeriesForm_deDonder}, followed by iterating the right hand side to any desired order in $\GN$; the $n$th order iteration would yield a solution accurate up to $\mathcal{O}[\GN^{n+1}]$. Since $\bar{h}_{\mu\nu}$ begins at first order in $\GN$, we note that $\delta_n t_{\mu\nu} = -\delta_n G_{\mu\nu}/(8\pi\GN)$ must begin at order $\GN^{n-1}$. In this work, I will be content with obtaining the first order in $\GN$ -- i.e., the leading contribution from $\delta_2 t_{\mu\nu}$ -- and in the non-relativistic contributions to the energy and angular momentum flux. Our central goals are therefore to calculate the linearized solutions to $\bar{h}_{\alpha\beta}$ and, from them, the `on-shell' $n=2$ term of eq. \eqref{GR_GWtmunu} energy flux
\begin{align}
\label{GR_EnergyFlux}
\frac{\dd E}{\dd t \dd \Omega}
= \lim_{r \to \infty} r^{d-2} \delta_2 t^{0i} \widehat{r}^i
= -(8\pi\GN)^{-1} \lim_{r \to \infty} r^{d-2} \delta_2 G^{0i} \widehat{r}^i ;
\end{align}
and angular momentum flux
\begin{align}
\label{GR_AngularMomentumFlux}
\frac{\dd L^{ij}}{\dd t \dd \Omega}
= \lim_{r \to \infty} r^{d-2} x^{[i} \delta_2 t^{j]k} \widehat{r}^k
= -(8\pi\GN)^{-1} \lim_{r \to \infty} r^{d-2} x^{[i} \delta_2 G^{j]k} \widehat{r}^k 
\end{align}
of gravitational radiation. Notice, away from the matter source and when evaluating on the linearized solutions of eq. \eqref{GR_SeriesForm_deDonder} -- i.e., $\delta_1 G_{\mu\nu} = 8\pi\GN \bar{T}_{\mu\nu}$ -- the indices of $\delta_2 G_{\mu\nu}$ may be moved with the flat metric $\eta^{\mu\nu}$ because $\delta_1 G_{\mu\nu} = 0$ in vacuum. Hence, in the far zone,
{\allowdisplaybreaks\begin{align}
\label{EinsteinTensor_2ndOrder}
\delta_2 G_{\rho\sigma}
&= \left(\delta_\rho^\mu \delta_\sigma^\nu - \frac{1}{2} \eta_{\rho\sigma} \eta^{\mu\nu}\right) \delta_2 R_{\mu\nu} , \\
\delta_2 R_{\mu\nu}
&= \frac{1}{2} \Bigg\{ 
	\frac{1}{2} \partial_{\mu} h_{\alpha\beta} \partial_\nu h^{\alpha\beta}
	+ h^{\alpha\beta} \left( \partial_{\nu} \partial_{\mu} h_{\alpha\beta} + \partial_{\beta} \partial_{\alpha} h_{\mu\nu} - \partial_{\beta} \partial_{\nu} h_{\mu\alpha} - \partial_{\beta} \partial_{\mu} h_{\nu\alpha} \right)  \\
&\qquad\qquad
	+ \partial^{\beta} h^{\alpha}_{\phantom{\alpha}\nu} \left( \partial_{\beta} h_{\mu\alpha} - \partial_{\alpha} h_{\mu\beta} \right)
	- \partial_\beta \left( h^{\alpha\beta} - \frac{1}{2} \eta^{\alpha\beta} h \right)
	\left( \partial_{\{\nu} h_{\mu\}\alpha} - \partial_\alpha h_{\mu\nu} \right) 
	\Bigg\} , \nonumber
\end{align}}
and
\begin{align}
	\delta_2 G^{\rho\sigma} = \eta^{\rho\mu} \eta^{\sigma\nu} \delta_2 G_{\mu\nu} .
\end{align}
{\bf Linearized Einstein's Equations} \qquad As already alluded to, the first order in $\GN$ solutions to $\bar{h}_{\alpha\beta}$ satisfy the linearized version of Einstein's equations in eq. \eqref{GR_SeriesForm_deDonder}:
\begin{align}
\partial^2 \bar{h}_{\mu\nu} = -16\pi \GN \bar{T}_{\mu\nu} .
\end{align}
Comparison with eq. \eqref{MasslessScalar_WaveEqn} immediately tells us, we have at hand a $d \times d$ matrix of massless scalar wave equations. Its solution can therefore be obtained by simply replacing in \S \eqref{Section_GreenFarZone} all the $\psi$s with $\bar{h}_{\mu\nu}$ and all the $J$s with $-16\pi\GN \bar{T}_{\mu\nu}$. With this in mind, let us turn to its far zone solutions needed for equations \eqref{GR_EnergyFlux} and \eqref{GR_AngularMomentumFlux}.

{\it Far Zone Limits} \qquad In the $r \to \infty$ limits and in even dimensions $d=4+2n$; with the relativistic retarded time denoted as $\tau \equiv t-r+\widehat{r} \cdot \vec{x}'$, the zeroth, first, and second derivatives are respectively
{\allowdisplaybreaks\begin{align}
		\label{GR_Linearized_EvenDim_0D}
		\bar{h}_{\mu\nu}[t,\vec{x}]
		&= -\frac{16\pi \GN}{2(2\pi r)^{1+n}} \int_{\mathbb{R}^{d-1}} \dd^{d-1}\vec{x}' \Bigg( \partial_t^n \bar{T}_{\mu\nu}[\tau,\vec{x}'] \\
		&\qquad\qquad
		+ \frac{1}{2} \frac{n(n+1) + (\widehat{r} \cdot \vec{x}') (2n+2) \partial_t - (r'^2-(\widehat{r} \cdot \vec{x}')^2) \partial_t^2}{r} \partial_t^{n-1} \bar{T}_{\mu\nu}[\tau,\vec{x}'] + \mathcal{O}[r^{-2}] \Bigg) , \nonumber
\end{align}}
{\allowdisplaybreaks\begin{align}
		\label{GR_Linearized_EvenDim_1D}
		\partial_{\alpha} \bar{h}_{\mu\nu}[t,\vec{x}]
		&= -\frac{16\pi \GN}{2(2\pi r)^{1+n}} \int_{\mathbb{R}^{d-1}} \dd^{d-1}\vec{x}' \Bigg( \left(\delta_{\alpha}^0 - \delta_{\alpha}^l \widehat{r}^l\right) \partial_t^{n+1} \bar{T}_{\mu\nu}[\tau,\vec{x}'] \\
		&+ \delta_{\alpha}^a P^{ab} \frac{x'^b}{r} \partial_t^{n+1} \bar{T}_{\mu\nu}[\tau,\vec{x}']
		-\frac{n+1}{r} \delta_{\alpha}^l \widehat{r}^l \partial_t^n \bar{T}_{\mu\nu}[\tau,\vec{x}'] \nonumber\\
		&+ \frac{1}{2} \frac{n(n+1) + (\widehat{r} \cdot \vec{x}') (2n+2) \partial_t - (r'^2-(\widehat{r} \cdot \vec{x}')^2) \partial_t^2}{r} \left(\delta_{\alpha}^0 - \delta_{\alpha}^l \widehat{r}^l\right) \partial_t^{n} \bar{T}_{\mu\nu}[\tau,\vec{x}'] + \mathcal{O}[r^{-2}] \Bigg) , \nonumber
\end{align}}
and
{\allowdisplaybreaks\begin{align}
		\label{GR_Linearized_EvenDim_2D}
		&\partial_{\alpha} \partial_\beta \bar{h}_{\mu\nu}[t,\vec{x}] \\
		&= -\frac{16\pi \GN}{2(2\pi r)^{1+n}} \int_{\mathbb{R}^{d-1}} \dd^{d-1}\vec{x}' \Bigg( \left(\delta_{\alpha}^0 - \delta_{\alpha}^l \widehat{r}^l\right) \left(\delta_{\beta}^0 - \delta_{\beta}^m \widehat{r}^m\right) \partial_t^{n+2} \bar{T}_{\mu\nu}[\tau,\vec{x}'] 
		- \delta_{\alpha}^l \frac{P^{lm}}{r} \delta^m_\beta \partial_t^{n+1} \bar{T}_{\mu\nu}[\tau,\vec{x}'] \nonumber\\
		&+ \left(\delta_{\{\alpha}^0 - \delta_{\{\alpha}^l \widehat{r}^l\right) \delta_{\beta\}}^a P^{ab} \frac{x'^b}{r} \partial_t^{n+2} \bar{T}_{\mu\nu}[\tau,\vec{x}'] 
		- \frac{n+1}{r} \delta_{\{\alpha}^l \widehat{r}^l \left( \delta_{\beta\}}^0 - \delta_{\beta\}}^m \widehat{r}^m \right) \partial_t^{n+1} \bar{T}_{\mu\nu}[\tau,\vec{x}'] \nonumber\\
		&+ \frac{1}{2} \frac{n(n+1) + (\widehat{r} \cdot \vec{x}') (2n+2) \partial_t - (r'^2-(\widehat{r} \cdot \vec{x}')^2) \partial_t^2}{r} \left( \delta_{\alpha}^0 - \delta_{\alpha}^l \widehat{r}^l \right) \left( \delta_{\beta}^0 - \delta_{\beta}^m \widehat{r}^m \right) \partial_t^{n+1} \bar{T}_{\mu\nu}[\tau,\vec{x}'] + \mathcal{O}[r^{-2}] \Bigg) . \nonumber
\end{align}}
In the same far zone limit but odd dimensional $d=3+2n$ flat background spacetimes; with the relativistic retarded time now involving an additional history integral $\tau \equiv t-r+\widehat{r} \cdot \vec{x}' - \mu$, the zeroth, first, second derivatives are respectively
\begin{align}
	\label{GR_Linearized_OddDim_0D}
	\bar{h}_{\mu\nu}[t,\vec{x}]
	&= -\frac{16 \pi \GN}{\sqrt{2} (2\pi)^{n+1} \cdot r^{n+\frac{1}{2}}} \int_{\mathbb{R}^{d-1}} \dd^{d-1}\vec{x}' \int_{0}^{\infty} \dd\mu \exp\left[ - \mu \cdot 0^+ \right]
	\Bigg( \mu^{-\frac{1}{2}} \partial_\tau^n \bar{T}_{\mu\nu}\left[ \tau,\vec{x}' \right] \\
	&+ \frac{\mu^{\frac{1}{2}}}{r} \left( \left(n^2 - \frac{1}{4}\right) \partial_\tau^n \bar{T}_{\mu\nu}\left[ \tau,\vec{x}' \right] + (\widehat{r} \cdot \vec{x}') (2n+1) \partial_\tau^{n+1} \bar{T}_{\mu\nu}\left[ \tau,\vec{x}' \right] - (r'^2-(\widehat{r}\cdot\vec{x}')^2) \partial_\tau^{n+2} \bar{T}_{\mu\nu}\left[ \tau,\vec{x}' \right] \right)
	+ \mathcal{O}[r^{-2}] \Bigg) , \nonumber
\end{align}
{\allowdisplaybreaks\begin{align}
\label{GR_Linearized_OddDim_1D}
\partial_\alpha \bar{h}_{\mu\nu}
&= -\frac{16 \pi \GN}{\sqrt{2} (2\pi)^{n+1} \cdot r^{n+\frac{1}{2}}} \int_{\mathbb{R}^{d-1}} \dd^{d-1}\vec{x}' \int_{0}^{\infty} \dd\mu \exp\left[ - \mu \cdot 0^+ \right] \\
&\times 
\Bigg\{ \mu^{-\frac{1}{2}} \left( \delta_{\alpha}^0 - \delta_{\alpha}^j \widehat{r}^j \right) \partial_\tau^{n+1} \bar{T}_{\mu\nu}\left[ \tau,\vec{x}' \right] + \frac{\mu^{-\frac{1}{2}}}{r} \left( \delta_{\alpha}^{a} P^{ab} x'^b \partial_\tau^{n+1} \bar{T}_{\mu\nu}\left[ \tau,\vec{x}' \right]  - \left( n + \frac{1}{2} \right) \widehat{r}^l \delta_{\alpha}^l \partial_\tau^n \bar{T}_{\mu\nu}\left[ \tau,\vec{x}' \right] \right) \nonumber\\
&\qquad\qquad
+ \frac{\mu^{-\frac{1}{2}}}{2r} \left( \delta_{\alpha}^0 - \delta_{\alpha}^j \widehat{r}^j \right) \Bigg( \left(n^2 - \frac{1}{4}\right) \partial_\tau^{n} \bar{T}_{\mu\nu}\left[ \tau,\vec{x}' \right] \nonumber\\
&\qquad\qquad\qquad\qquad
+ (\widehat{r} \cdot \vec{x}') (2n+1) \partial_\tau^{n+1} \bar{T}_{\mu\nu}\left[ \tau,\vec{x}' \right] - (r'^2 - (\widehat{r}\cdot\vec{x}')^2) \partial_\tau^{n+2} \bar{T}_{\mu\nu}\left[ \tau,\vec{x}' \right] \Bigg)  
+ \mathcal{O}[r^{-2}] \Bigg\} \nonumber
\end{align}}
and
{\allowdisplaybreaks\begin{align}
\label{GR_Linearized_OddDim_2D}
&\partial_\alpha \partial_\beta \bar{h}_{\mu\nu}[t,\vec{x}] \\
&= -\frac{16 \pi \GN}{\sqrt{2} (2\pi)^{n+1} \cdot r^{n+\frac{1}{2}}} \int_{\mathbb{R}^{d-1}} \dd^{d-1}\vec{x}' \int_{0}^{\infty} \dd\mu \exp\left[ - \mu \cdot 0^+ \right] \nonumber\\
&\times 
\Bigg\{ \mu^{-\frac{1}{2}} \left( \delta_{\alpha}^0 - \delta_{\alpha}^j \widehat{r}^j \right) \left( \delta_{\beta}^0 - \delta_{\beta}^k \widehat{r}^k \right) \partial_\tau^{n+2} \bar{T}_{\mu\nu}\left[ \tau,\vec{x}' \right] 
\nonumber\\
&+ \frac{\mu^{-\frac{1}{2}}}{r} \Bigg\{ \left( \delta_{\{\alpha}^0 - \delta_{\{\alpha}^j \widehat{r}^j \right) \left( \delta_{\beta\}}^{a} P^{ab} x'^b \partial_\tau^{n+2} \bar{T}_{\mu\nu}\left[ \tau,\vec{x}' \right]  - \left( n + \frac{1}{2} \right) \widehat{r}^l \delta_{\beta\}}^l \partial_\tau^{n+1} \bar{T}_{\mu\nu}\left[ \tau,\vec{x}' \right] \right)
- \delta_\alpha^a P^{ab} \delta_\beta^b \partial_\tau^{n+1} \bar{T}_{\mu\nu}\left[ \tau, \vec{x}' \right]  \Bigg\} \nonumber\\
&+ \frac{\mu^{-\frac{1}{2}}}{2r} \left( \delta_{\alpha}^0 - \delta_{\alpha}^j \widehat{r}^j \right) \left( \delta_{\beta}^0 - \delta_{\beta}^k \widehat{r}^k \right) \Bigg( \left(n^2 - \frac{1}{4}\right) \partial_\tau^{n+1} \bar{T}_{\mu\nu}\left[ \tau,\vec{x}' \right] \nonumber\\
&\qquad\qquad\qquad\qquad
+ (\widehat{r} \cdot \vec{x}') (2n+1) \partial_\tau^{n+2} \bar{T}_{\mu\nu}\left[ \tau,\vec{x}' \right] - (r'^2 - (\widehat{r}\cdot\vec{x}')^2) \partial_\tau^{n+3} \bar{T}_{\mu\nu}\left[ \tau,\vec{x}' \right] \Bigg) + \mathcal{O}[r^{-2}] \Bigg\} . \nonumber
\end{align}}
{\bf Non-relativistic Limit} \qquad I now turn to the non-relativistic limit by first assuming that the momentum and shear-stress of matter scales respectively as $v$ and $v^2$ relative to the energy density, namely $T_{0i} \sim v \cdot T_{00}$ and $T_{ij} \sim v^2 \cdot T_{00}$, where $v \ll 1$ is some characteristic speed of its internal dynamics. Furthermore, I will only study the case where the contribution to the total energy-momentum-shear-stress is dominated by that of the matter itself, so that 
\begin{align}
\label{ConservationOfTmunu}
\partial^\mu \bar{T}_{\mu\nu} = \partial_0 \bar{T}_{0\nu} - \partial_i \bar{T}_{i\nu} = 0 .
\end{align}
This unfortunately {\it does not} cover the physically important case of the compact binary system in 4D -- nor any system bound by its self-gravity. For, in such a scenario both gravitational and matter necessarily contribute significantly to the system's total energy-momentum tensor. In fact, because eq. \eqref{ConservationOfTmunu} does not contain any coupling to gravity, if it were to hold for the compact binary system, it would imply the two objects would be traveling on independent straight lines in otherwise empty space -- namely, a bound system cannot exist.

By taking the time derivative of the total mass and spatial momentum
\begin{align}
	\label{GR_Mass}
	M 	&\equiv \int_{\mathbb{R}^{d-1}} \bar{T}^{00}[t,\vec{x}'] \dd^{d-1}\vec{x}', \\
	\label{GR_Momentum}
	P^i	&\equiv \int_{\mathbb{R}^{d-1}} \bar{T}^{0i}[t,\vec{x}'] \dd^{d-1}\vec{x}';
\end{align}
followed by using eq. \eqref{ConservationOfTmunu}, one may learn that $M$ and $P^i$ are time-independent for a physically isolated body. Along similar lines, the dipole moment -- proportional to the leading order center-of-mass --
\begin{align}
	\label{GR_Dipole}
	d^i[t] \equiv \int_{\mathbb{R}^{d-1}} x'^i \bar{T}_{00}[t,\vec{x}'] \dd^{d-1}\vec{x}',
\end{align}
obeys
\begin{align}
	\label{GR_DipoleVelocityIsMomentum}
	\partial_t d^i[t] = P^i .
\end{align}
Below, we will see that gravitational energy-momentum radiation begins at the quadrupole order in all dimensions because the matter energy-momentum $(M,P^i)$ is conserved

Next, the double time derivative of the quadrupole moment
\begin{align}
	\label{GR_Quadrupole}
	Q_{ij}[t] \equiv \int_{\mathbb{R}^{d-1}} x'^i x'^j \bar{T}^{00}[t,\vec{x}'] \dd^{d-1}\vec{x}'
\end{align}
is related to the spatial components of the energy momentum tensor via the relation
\begin{align}
	\label{GR_Quadrupole_Tij}
	\frac{1}{2} \partial_t^2 Q_{ij}[t] = \int_{\mathbb{R}^{d-1}} \bar{T}_{ij}[t,\vec{x}'] \dd^{d-1}\vec{x}' .
\end{align}
We may now write the de Donder gauge $\bar{h}_{\mu\nu}$ solutions up to the quadrupole order in the non-relativistic expansion. Similar to the electromagnetic case, this is achieved by Taylor expanding the stress tensors in equations \eqref{GR_Linearized_EvenDim_0D} through \eqref{GR_Linearized_OddDim_2D} with respect to time, in powers of $\widehat{r} \cdot \vec{x}'$; as well as utilizing equations \eqref{GR_Mass}--\eqref{GR_Quadrupole_Tij} throughout these calculations. At leading order, the $\bar{h}_{\mu\nu}$ and its first and second derivatives are built entirely out of the monopole $M$, dipole $\vec{d}$, and quadrupole moment $Q_{ij}$. For instance, in even dimensions $d=4+2n$, the $00$ of the de Donder gauge metric perturbations now takes the form 
{\allowdisplaybreaks\begin{align}
\label{hb00_EvenDim_NR}
\bar{h}_{00}[t,\vec{x}] 
		&\approx -\frac{16\pi \GN}{2(2\pi r)^{1+n}} \Bigg( \partial_t^n M + \widehat{r} \cdot \partial_t^n \vec{P} + \frac{1}{2} \widehat{r}^a \widehat{r}^b \partial_t^{n+2} Q_{ab}[t-r]  \\
		&+ \frac{1}{r} \Bigg\{ \frac{n(n+1)}{2} \partial_t^{n-1} M + \frac{(n+1)(n+2)}{2} \widehat{r} \cdot \partial_t^n \vec{d}[t-r]
		- \left(\frac{\delta^{ab}}{2} - \frac{(n+2)(n+3)}{4} \widehat{r}^a \widehat{r}^b\right) \partial_t^{n+1} Q_{ab}[t-r] \Bigg\} \Bigg) . \nonumber
\end{align}}
Whereas the spatial vector part becomes
{\allowdisplaybreaks\begin{align}
\label{hb0i_EvenDim_NR}
\bar{h}_{0i}[t,\vec{x}]
		&\approx -\frac{16\pi \GN}{2(2\pi r)^{1+n}} \Bigg( \partial_t^n P_i - \frac{\widehat{r}^a}{2} \partial_t^{n+2} Q_{ai}[t-r] + \dots \\
		&\qquad\qquad
		+ \frac{1}{r} \left\{ \frac{n(n+1)}{2} \partial_t^{n-1} P_i - \frac{(n+1)(n+2)}{4} \widehat{r}^a \partial_t^{n+1} Q_{ai}[t-r] + (n+1) \xi_i[t_0] \right\} \Bigg) , \nonumber
\end{align}}
where, for some arbitrary time $t_0$, the $\xi_i$ is defined through the relations
\begin{align}
\label{xi_ODE_Soln}
\int_{\mathbb{R}^{d-1}} \dd^{d-1}\vec{x}' (\widehat{r} \cdot \vec{x}') \bar{T}_{0i}[t,\vec{x}']
	&= -\frac{\widehat{r}^a}{2} \partial_t Q_{ai}[t] + \xi_i[t_0] \\
\label{xi_ODE_BC}
\xi_i[t_0]
	&\equiv \frac{\widehat{r}^a}{2} \partial_t Q_{ai}[t_0] 
	+ \int_{\mathbb{R}^{d-1}} \dd^{d-1}\vec{x}' (\widehat{r} \cdot \vec{x}') \bar{T}_{0i}[t_0,\vec{x}'] .
\end{align}
That eq. \eqref{xi_ODE_Soln} holds can be seen by verifying -- via the conservation law in eq. \eqref{ConservationOfTmunu} and the relationship between the quadrupole's acceleration and the shear-stress of eq. \eqref{GR_Quadrupole_Tij} -- the (first order) ordinary differential equation
\begin{align}
\frac{\dd}{\dd t}\int_{\mathbb{R}^{d-1}} \dd^{d-1}\vec{x}' (\widehat{r} \cdot \vec{x}') \bar{T}_{0i}[t,\vec{x}']
	&= -\frac{\widehat{r}^a}{2} \partial_\tau^2 Q_{ai}[t] .
\end{align}
The $\xi_i$ in eq. \eqref{xi_ODE_BC} are therefore the `initial conditions'.

Finally, the rank-2 spatial components of the metric perturbations are now
{\allowdisplaybreaks\begin{align}
\label{hbij_EvenDim_NR}
\bar{h}_{ij}[t,\vec{x}]
		&\approx -\frac{16\pi \GN}{2(2\pi r)^{1+n}} \left(\frac{1}{2} \partial_t^{n+2} Q_{ij}[t-r] + \frac{n(n+1)}{4r} \partial_t^{n+1} Q_{ij}[t-r] \right).
\end{align}}
Turning to the odd dimensional $d=3+2n$ case, the metric perturbations now feature integrals over the retarded history of the quadrupole moments.
{\allowdisplaybreaks\begin{align}
\label{hb00_OddDim_NR}
\bar{h}_{00}[t,\vec{x}]
&\approx -\frac{16 \pi \GN}{\sqrt{2} (2\pi)^{n+1} \cdot r^{n+\frac{1}{2}}} \Bigg( \int_{0}^{\infty} e^{-\mu\cdot 0^+} \left(\partial_t^n M + \widehat{r} \cdot \partial_t^n \vec{P} + \frac{\widehat{r}^a \widehat{r}^b}{2} \partial_t^{n+2} Q_{ab}[t-r-\mu]\right) \frac{\dd\mu}{\sqrt{\mu}} \\
&+\frac{1}{r} \Bigg\{ \frac{2n+1}{4} \int_{0}^{\infty} e^{-\mu\cdot 0^+} \left(3\widehat{r}\cdot\partial_t^n\vec{P} - \partial_t^n M \right) \sqrt{\mu}\dd\mu \nonumber\\
&\qquad\qquad\qquad
+ \int_{0}^{\infty} e^{-\mu\cdot 0^+} \left(\frac{(2n+3)(2n+5)}{8}\widehat{r}^a\widehat{r}^b - \delta^{ab}\right) \partial_t^{n+1} Q_{ab}[t-r-\mu] \frac{\dd\mu}{2\sqrt{\mu}}
 \Bigg\} \Bigg) \nonumber\\
\label{hb0i_OddDim_NR}
\bar{h}_{0i}[t,\vec{x}]
&\approx -\frac{16 \pi \GN}{\sqrt{2} (2\pi)^{n+1} \cdot r^{n+\frac{1}{2}}} \Bigg( \int_{0}^{\infty} e^{-\mu\cdot 0^+} \left( \partial_t^n P_i - \frac{\widehat{r}^a}{2} \partial_t^{n+2} Q_{ai}[t-r-\mu] \right) \frac{\dd\mu}{\sqrt{\mu}} \\
&+ \frac{1}{r} \left\{ \frac{4n^2-1}{4} \int_{0}^{\infty} e^{-\mu\cdot 0^+} \partial_t^n P_i \sqrt{\mu}\dd\mu 
- \frac{(2n+1)(2n+3)}{8} \int_{0}^{\infty} e^{-\mu\cdot 0^+} \widehat{r}^a \partial_t^{n+1} Q_{ai}[t-r-\mu] \frac{\dd\mu}{2\sqrt{\mu}}
\right\} \Bigg) \nonumber\\
\label{hbij_OddDim_NR}
\bar{h}_{ij}[t,\vec{x}]
&\approx -\frac{16 \pi \GN}{\sqrt{2} (2\pi)^{n+1} \cdot r^{n+\frac{1}{2}}} \int_{0}^{\infty} e^{-\mu\cdot 0^+} \left( \frac{1}{2} \partial_t^{n+2} Q_{ij}[t-r-\mu]
+ \frac{4n^2-1}{16 \cdot r} \partial_t^{n+1} Q_{ij}[t-r-\mu] \right) \frac{\dd\mu}{\sqrt{\mu}} 
\end{align}}
In equations \eqref{hb00_OddDim_NR}, \eqref{hb0i_OddDim_NR} and \eqref{hbij_OddDim_NR}, we have converted one of the time derivatives acting on the quadrupole occurring within the $\int_{0}^{\infty} \dots \sqrt{\mu} \dd\mu$ integral into a negative derivative with respect to $\mu$, followed by integrating it by parts. (As explained before, the upper limit of the boundary term is zero due to the $e^{-\mu\cdot 0^+}$ and the lower limit is zero due to the $\sqrt{\mu}$.) Furthermore, notice, for $n=0$ (i.e., $d=3$), the integrals involving $M$ and $\widehat{r} \cdot \vec{P}$ are divergent; whereas this infinity is absent for all higher odd dimensions. In fact, the de Donder gauge appears to be violated by the divergent $(1/r)\int_{0}^{\infty} \dots \sqrt{\mu} \dd\mu$ integrals in the second lines of equations \eqref{hb00_OddDim_NR} and \eqref{hb0i_OddDim_NR}; for, a direct computation will teach us
\begin{align}
\label{deDonderGauge_Violation}
\partial^\mu \bar{h}_{\mu\nu} 
&= -\frac{16 \pi \GN}{\sqrt{2} (2\pi)^{n+1} \cdot r^{n+\frac{1}{2}}} \cdot \frac{4 n^2 - 1}{8 r} \int_{0}^{\infty} e^{-\mu\cdot 0^+} \left( \delta_\nu^i \partial_t^n P_i + \delta_\nu^0 \partial_t^n M \right) \sqrt{\mu}\dd\mu .
\end{align}
This suggests the $d=3$ case requires special care to develop a proper asymptotic series.

I also highlight here, in a similar spirit to the electromagnetic case, it is important {\it not} to take the non-relativistic limit -- replacing $t-r+\widehat{r}\cdot\vec{x}'$ with $t-r$ -- too early. For example, in 4D, if we take from the outset $\bar{h}_{\mu\nu} = -(4\GN/r) \int \dd^3\vec{x}' \bar{T}_{\mu\nu}[t-r,\vec{x}']$, a quick calculation would reveal the de Donder gauge does not hold ($\partial^\mu \bar{h}_{\mu\nu} \neq 0$) and therefore eq. \eqref{ConservationOfTmunu} cannot be used consistently to re-write $\bar{h}_{\mu\nu}$ and its first and second derivatives in terms of quadrupole moments.

To reiterate: the solutions in equations \eqref{hb00_EvenDim_NR}, \eqref{hb0i_EvenDim_NR}, \eqref{hbij_EvenDim_NR}, \eqref{hb00_OddDim_NR}, \eqref{hb0i_OddDim_NR} and \eqref{hbij_OddDim_NR} are equations \eqref{GR_Linearized_EvenDim_0D} and \eqref{GR_Linearized_OddDim_0D} expanded up to the quadrupole order and up to $1/r^{d/2}$ order, with {\it relative} corrections that scale as $(\widehat{r} \cdot \vec{x}') \partial_t \sim v$ or $r_s/r$, where $r_s$ is the characteristic size of the matter source. I have also carried out this non-relativistic expansion for $\partial_\alpha \bar{h}_{\mu\nu}$ and $\partial_\alpha \partial_\beta \bar{h}_{\mu\nu}$. Furthermore, as a check of my non-relativistic calculations, I have verified that both the wave equation $\partial^2 \bar{h}_{\mu\nu}=0$ and the de Donder gauge $\partial^\mu \bar{h}_{\mu\nu} = 0 = \partial^\mu \partial_\alpha \bar{h}_{\mu\nu}$ are satisfied for all $d \geq 4$.

{\bf Gravitational Radiation: Energy Flux} \qquad For even dimensional $d=4+2n$ asymptotically flat spacetimes, inserting equations \eqref{hb00_EvenDim_NR}, \eqref{hb0i_EvenDim_NR}, \eqref{hbij_EvenDim_NR} and their first and second derivative results into equations \eqref{GR_EnergyFlux} and \eqref{EinsteinTensor_2ndOrder} -- together with considerable aid from {\sf xAct} \cite{xAct} -- reveals
\begin{align}
\label{GR_EnergyFlux_EvenDim_v1}
\delta_2 t_{\mu\nu}
&= -\frac{1}{8\pi\GN} \left(\frac{16\pi \GN}{2(2\pi r)^{1+n}}\right)^2 \cdot \partial_\mu (t-r) \partial_\nu \left(t-r\right) \\
&\qquad\qquad \times \left( -\frac{\partial_t^{n+3} Q^{\text{(tt)}}_{ab}[t-r] \partial_t^{n+3} Q^{\text{(tt)}}_{ab}[t-r]}{16} + \frac{\partial}{\partial t} \frac{\partial_t^{n+2} Q^{\text{(tt)}}_{ab}[t-r] \partial_t^{n+3} Q^{\text{(tt)}}_{ab}[t-r]}{8}\right)  \nonumber\\
&\qquad\qquad 
+ \text{terms linear in the quadrupole moments $\partial_t^{n+4} Q_{ab}[t-r]$} , \nonumber
\end{align}
where the `transverse-traceless' quadrupole is defined as
\begin{align}
\label{GR_Quadrupole_tt}
Q_{ab}^\text{(tt)} 
&\equiv P_{ab ij} Q_{ij} , \\
\label{GR_ttProjector}
P_{ab ij}
&\equiv \frac{1}{2} P_{a\{i} P_{j\}b} - \frac{P_{ab} P_{ij}}{d-2},
\qquad
P_{ij} \equiv \delta_{ij} - \widehat{r}_i \widehat{r}_j .
\end{align}
Due to the projector properties in eq. \eqref{Pab_Properties} as well as $P^{aa} = d-2$,
\begin{align}
\label{tt}
\delta^{ab} Q_{ab}^\text{(tt)} = 0 = \widehat{r}^a Q_{ab}^\text{(tt)} .
\end{align}
The second term of the second line of eq. \eqref{GR_EnergyFlux_EvenDim_v1} and the terms linear in quadrupole moments would contribute as a total time derivative to the gravitational energy flux $\dd E/(\dd t \dd \Omega)$ in eq. \eqref{GR_EnergyFlux}. If we assume
\begin{align}
\left. \partial_t^{n+2} Q_{ab}[t-r] \partial_t^{n+3} Q_{ij}[t-r] \right\vert_{t= \pm \infty} 
= 0 =
\left. \partial_t^{n+3} Q_{ab}[t-r] \right\vert_{t= \pm \infty} ,
\end{align}
then such a total time derivative would not contribute to the {\it total} energy radiated. Moreover, one may readily check that 
\begin{align}
\partial^\mu \left( \frac{\partial_\mu (t-r) \partial_\nu \left(t-r\right)}{r^{d-2}} A[t-r] B[t-r] \right) = 0 
\end{align}
for all dimensions $d \geq 3$ and for arbitrary functions $A$ and $B$ of the retarded time $t-r$. Altogether, that implies we may discard the total time derivative terms of eq. \eqref{GR_EnergyFlux_EvenDim_v1} and identify the remaining term as the effective conserved stress-energy tensor of gravitational radiation:
\begin{align}
	\label{GR_EnergyFlux_EvenDim_tmunu}
	r^{d-2} \delta_2 t_{\mu\nu}^{\text{(eff)}}[d=4+2n]
	&= \frac{\GN}{2^{2n+3}\pi^{2n+1}} \cdot \partial_\mu (t-r) \partial_\nu \left(t-r\right) \partial_t^{n+3} Q^{\text{(tt)}}_{ab}[t-r] \partial_t^{n+3} Q^{\text{(tt)}}_{ab}[t-r] .
\end{align}
It is worth pointing out, the coefficients of $\partial_t^{n+4} Q_{ab}[t-r]$ in eq. \eqref{GR_EnergyFlux_EvenDim_v1} are either $M$ or $P^i$. If the energy-momentum $(M,P^i)$ were not constant, therefore, these $\partial_t^{n+4} Q_{ab}[t-r]$ terms would then contribute to $\delta_2 t_{\mu\nu}^{\text{(eff)}}$ because they are no longer total time derivatives.

Inserting eq. \eqref{GR_EnergyFlux_EvenDim_tmunu} into eq. \eqref{GR_EnergyFlux} hands us the differential gravitational wave energy flux per solid angle,
\begin{align}
	\label{GR_EnergyFlux_EvenDim_PerOmega}
	\frac{\dd E}{\dd t \dd \Omega}
	&= \frac{\GN}{2^{d-1}\pi^{d-3}} \partial_t^{\frac{d+2}{2}} Q^{\text{(tt)}}_{ab}[t-r] \partial_t^{\frac{d+2}{2}} Q^{\text{(tt)}}_{ab}[t-r], \qquad (\text{Even $d \geq 4$}) .
\end{align}
Because of the projector properties of $P_{ij}$ in eq. \eqref{Pab_Properties}, $P_{ab mn} P_{mn ij} = P_{ab ij}$, and the total energy flux can be obtained from eq. \eqref{GR_EnergyFlux_EvenDim_PerOmega} as
\begin{align}
\frac{\dd E}{\dd t}
&= \frac{\GN}{2^{d-1}\pi^{d-3}} \partial_t^{\frac{d+2}{2}} Q_{ab}[t-r] \partial_t^{\frac{d+2}{2}} Q_{ij}[t-r] \cdot \int_{\mathbb{S}^{d-2}} \left( \frac{1}{2} P_{a\{i} P_{j\}b} - \frac{P_{ab} P_{ij}}{d-2} \right) \dd^{d-2}\Omega .
\end{align}
These solid angle integrals involve Kronecker deltas, $\widehat{r}^i \widehat{r}^j$ and $\widehat{r}^i \widehat{r}^j \widehat{r}^a \widehat{r}^b$. Upon utilizing equations \eqref{SolidAngle}, \eqref{TensorIntegral_2PointFunction}, and \eqref{TensorIntegral_4PointFunction} of \S \eqref{Section_Calculus} below, we arrive at eq. \eqref{GR_EnergyFlux_EvenDim}.

Let us move on to the odd dimensional case $d=3+2n$. Using {\sf xAct} \cite{xAct} to plug equations \eqref{hb00_OddDim_NR}, \eqref{hb0i_OddDim_NR} and \eqref{hbij_OddDim_NR} into equations \eqref{GR_EnergyFlux} and \eqref{EinsteinTensor_2ndOrder}
\begin{align}
	\label{GR_EnergyFlux_OddDim_v1}
	\delta_2 t_{\mu\nu}
	&= -\frac{1}{8\pi\GN} \left(\frac{16 \pi \GN}{\sqrt{2} (2\pi)^{n+1} r^{n+\frac{1}{2}}}\right)^2 \cdot \partial_\mu (t-r) \partial_\nu \left(t-r\right) \\
	&\qquad\qquad \times \left( -\frac{\partial_t^{n+3} Q^{\text{(tt)}}_{ab}[t-r] \partial_t^{n+3} Q^{\text{(tt)}}_{ab}[t-r]}{16} + \frac{\partial}{\partial t} \frac{\partial_t^{n+2} Q^{\text{(tt)}}_{ab}[t-r] \partial_t^{n+3} Q^{\text{(tt)}}_{ab}[t-r]}{8}\right) , \nonumber\\
	&\qquad\qquad 
	+ \text{terms linear in the quadrupole moments $\int_{0}^{\infty} \frac{\dd\mu}{\sqrt{\mu}} \partial_t^{n+4} Q_{ab}[t-r-\mu]$} , \nonumber\\
Q_{ab}^{(tt)}[t-r]
&= P_{ab ij} \int_{0}^{\infty} \frac{\dd\mu}{\sqrt{\mu}} Q_{ij}[t-r-\mu] .
\end{align}
Note that the transverse and traceless properties of eq. \eqref{tt} render $Q_{ab}^{\text{(tt)}}$ trivial in 3D, i.e., for $n=0$. To see this, simply pick $\widehat{r}$ to point, say, along the $1-$axis. That $Q_{ab}^{\text{(tt)}}$ is transverse then implies $Q_{1a}^{\text{(tt)}} = 0 = Q_{a1}^{\text{(tt)}}$. The sole remaining non-trivial component $Q_{22}^{\text{(tt)}}$ must too be zero by the traceless condition. Hence, only the terms linear in the quadrupole moments survive. If this result holds up after a more careful asymptotic analysis is carried out that not only avoids the de Donder gauge breaking terms of eq. \eqref{deDonderGauge_Violation} but also the divergences involving $M$ and $P_i$, then it would confirm the widely accepted view that gravitational radiation does not exist in a 2+1 dimensional background Minkowski spacetime.

For all odd $d \geq 5$, as with the analysis in even dimensions, we discard all total time derivative terms. To this end I assume
\begin{align}
\left. \partial_t^{n+3} Q_{ab}[t-r-\mu] \right\vert_{t=-\infty}
= 0 =
\lim_{t\to\infty} \int_{-t}^{\infty} \frac{\dd\mu'}{\sqrt{\mu'+t}} e^{-(\mu'+t)\cdot 0^+} \partial_t^{n+3} Q_{ab}[-r-\mu'] ,
\end{align}
and thus only retain the first term of eq. \eqref{GR_EnergyFlux_OddDim_v1}. Following this I employ the resulting effective (conserved) stress tensor in eq. \eqref{GR_EnergyFlux}.
{\allowdisplaybreaks\begin{align}
	\label{GR_EnergyFlux_OddDim_tmunu}
	r^{d-2} \delta_2 t_{\mu\nu}^{\text{(eff)}}[d = 3+2n \geq 5]
	&= \frac{\GN}{2^{2n+2}\pi^{2n+1}} \cdot \partial_\mu (t-r) \partial_\nu \left(t-r\right) \partial_t^{n+3} Q^{\text{(tt)}}_{ab}[t-r] \partial_t^{n+3} Q^{\text{(tt)}}_{ab}[t-r] , \\
	\label{GR_EnergyFlux_OddDim_PerOmega}
	\frac{\dd E}{\dd t \dd \Omega}
&= \frac{\GN}{2^{d-1}\pi^{d-2}} \partial_t^{\frac{d+3}{2}} Q^{\text{(tt)}}_{ab}[t-r] \partial_t^{\frac{d+3}{2}} Q^{\text{(tt)}}_{ab}[t-r], \qquad (\text{Odd $d \geq 5$}) .
\end{align}}
Once again, we may recall equations \eqref{SolidAngle}, \eqref{TensorIntegral_2PointFunction}, and \eqref{TensorIntegral_4PointFunction} to integrate eq. \eqref{GR_EnergyFlux_OddDim_PerOmega} over the $(d-2)-$sphere to yield the total gravitational energy flux in eq. \eqref{GR_EnergyFlux_OddDim}. Just like the even dimensional case, let us also record here that the coefficients of the integral of $\partial_t^{n+4} Q_{ab}$ in eq. \eqref{GR_EnergyFlux_OddDim_v1} are either $M$ or $P^i$. If the energy-momentum $(M,P^i)$ were not constant, therefore, these $\partial_t^{n+4} Q_{ab}$ terms would then contribute to $\delta_2 t_{\mu\nu}^{\text{(eff)}}$ because they are no longer total time derivatives. Altogether, as already alluded to, we conclude that gravitational energy-momentum radiation begins at the quadrupole order because of matter energy-momentum conservation.

At this point, eq. \eqref{GR_EnergyFlux_FreqSpace} -- the frequency space counterpart to equations \eqref{GR_EnergyFlux_EvenDim} and \eqref{GR_EnergyFlux_OddDim} -- may be deduced from Parseval's theorem and the decomposition in odd dimensions
\begin{align}
	\int_{0}^{\infty} \frac{\dd \mu}{\sqrt{\mu}} e^{-\mu \cdot 0^+} Q_{ab}[t-r-\mu]
	&= \int \frac{\dd\omega}{2\pi} \frac{\sqrt{\pi}}{(-i (\omega + i 0^+))^{\frac{1}{2}}} \exp\left[-i \omega(t-r)\right] \widetilde{Q}_{ab}[\omega] .
\end{align}
As was the case for electromagnetism, the gravitational energy flux per angular frequency takes the same form for all $d \geq 4$, despite the distinction in the causal structure of the real-time signals in even versus odd dimensions.

{\bf Gravitational Radiation: Angular Momentum Flux} \qquad The gravitational angular momentum flux requires developing $\delta_2 t^{\mu\nu}$ to one higher order in the $1/r$ expansion, because of the extra factor of $r$ in the $x^{[i} T^{j]k} \widehat{r}^k$.

For $d=4+2n$, inserting equations \eqref{hb00_EvenDim_NR}, \eqref{hb0i_EvenDim_NR}, \eqref{hbij_EvenDim_NR} and their first and second derivative results into equations \eqref{GR_AngularMomentumFlux} and \eqref{EinsteinTensor_2ndOrder} using {\sf xAct} \cite{xAct}; discarding in the resulting expressions total derivative terms, which all are proportional to either
\begin{align}
\partial_t \left(\partial_t^{n+1} Q_{ab}[t-r] \partial_t^{n+3} Q_{ij}[t-r]\right),
\qquad
\partial_t^{n+3} Q_{ij}[t-r],
\qquad
\text{or}
\qquad
\partial_t^{n+4} Q_{ij}[t-r] ;
\end{align}
followed by integrating over the solid angle with equations \eqref{SolidAngle}, \eqref{TensorIntegral_2PointFunction} and \eqref{TensorIntegral_4PointFunction} below:
\begin{align}
	\label{GR_AngularMomentumFlux_EvenDim}
	\frac{\dd L^{ij}}{\dd t} 
	&= \frac{d(d-1)(d-3) \GN}{2^{d-1} (d-2) \pi^{\frac{d-5}{2}} \Gamma[\frac{d+3}{2}]} \partial_t^{\frac{d}{2}} Q_{a[i}[t-r] \partial_t^{\frac{d+2}{2}} Q_{j]a}[t-r] 
	\qquad \text{(Even $d \geq 4$)}
\end{align}
The $d=4$ result recovers the spatial Hodge dual of eq. (4.9) of Peters \cite{Peters:1964zz}.
\begin{align}
\left. \frac{\dd L^{ij}}{\dd t} \right\vert_{d=4} 
	&= \frac{2}{5} \GN \partial_t^{2} Q_{a[i}[t-r] \partial_t^{3} Q_{j]a}[t-r] .
\end{align}
For the odd dimensional case, $d=3+2n$, using {\sf xAct} \cite{xAct} to plug equations \eqref{hb00_OddDim_NR}, \eqref{hb0i_OddDim_NR} and \eqref{hbij_OddDim_NR} into equations \eqref{GR_AngularMomentumFlux} and \eqref{EinsteinTensor_2ndOrder}; dropping in the ensuing expressions all total time derivative terms -- which are all proportional to
\begin{align}
\frac{\partial}{\partial t} \left( \int_{0}^{\infty} e^{-\mu \cdot 0^+} \partial_t^{n+3} Q_{ab}^{n+3}[t-r-\mu] \frac{\dd\mu}{\sqrt{\mu}} \cdot \int_{0}^{\infty} e^{-\mu' \cdot 0^+} \partial_t^{n+1} Q_{ab}[t-r-\mu'] \frac{\dd\mu'}{\sqrt{\mu'}}
\right) ;
\end{align}
followed by integrating over the solid angle: 
\begin{align}
	\label{GR_AngularMomentumFlux_OddDim}
	\frac{\dd L^{ij}}{\dd t} 
	&= \frac{d(d-1)(d-3) \GN}{2^{d-1} (d-2) \pi^{\frac{d-3}{2}} \Gamma[\frac{d+3}{2}]} \int_{0}^{\infty} \partial_t^{\frac{d+1}{2}} Q_{a[i}[t-r-\mu] \frac{\dd\mu}{\sqrt{\mu}} \cdot \int_{0}^{\infty} \partial_t^{\frac{d+3}{2}} Q_{j]a}[t-r-\mu'] \frac{\dd\mu'}{\sqrt{\mu'}} .
\end{align}
In frequency space,
\begin{align}
	\frac{\dd L^{ij}}{\dd\omega} 
	&= \frac{d(d-1)(d-3) \GN}{2^{d} (d-2) \pi^{\frac{d-3}{2}} \Gamma[\frac{d+3}{2}]} \omega^{d+1} \widetilde{Q}_{a[i}[\omega] \widetilde{Q}_{j]a}[\omega]^* .
\end{align}
Once again, the angular momentum flux per angular frequency takes the same form in all dimensions $d \geq 4$.

{\bf Finite Duration Quasi-Periodic Quadrupole Source} \qquad I close this section with a brief discussion to parallel the one at the end of \S \eqref{Section_EM}, by considering a finite duration (quasi-)periodic quadrupole source of gravitational waves. Instead of studying the real-time energy flux, however, I will instead study the synchronous gauge metric perturbation $h_{ij}^{(s)}$ to leading order in $1/r^{(d/2)-1}$. This is because $h_{ij}^{(s)}$ is directly linked to the fractional distortion of space itself, which is responsible for the changes in the interference patterns measured by laser interferometer based gravitational wave detector. Specifically, if the initial perturbations are zero, then the fractional distortion $\delta L/L$ between the spatial locations $\vec{X}$ and $\vec{Y}$, whose associated unit direction vector I dub here as $\widehat{n} \equiv (\vec{X}-\vec{Y})/|\vec{X}-\vec{Y}|$, is given by the formula\footnote{See, for example, Appendix C of \cite{Chu:2016ngc}.}
\begin{align}
\label{FractionalDistortion}
\frac{\delta L}{L} 
= -\frac{\widehat{n}^i \widehat{n}^j}{2} \int_{0}^{1} h^{(s)}_{ij}\left[t,\vec{X}+\lambda(\vec{Y}-\vec{X})\right] \dd\lambda .
\end{align}
I remark in passing that it is conceptually useful to relate $h^{(s)}_{ij}$ to the gauge-invariant linearized components of the Riemann tensor via
\begin{align}
R_{0i0j} = - \frac{1}{2} \partial_t^2 h^{(s)}_{ij}[t,\vec{x}] ;
\end{align}
so that eq. \eqref{FractionalDistortion} can in turn be re-cast in a manifestly gauge-invariant form, at the cost of introducing two extra integrals.

In the far zone $\omega r \gg 1$, the synchronous gauge metric perturbation is in fact equal to -- up to additive zero frequency initial conditions -- the transverse-traceless portion of the de Donder gauge spatial perturbation:\footnote{See eq. (46) of \cite{Chu:2019ndv}. The $h_{ij}^{(s)}$ and $h_{ij}^{(tt)}$ here are the $\chi_{ij}^{(\text{Synch})}$ and $\chi_{ij}^{(\text{tt})}$ in \cite{Chu:2019ndv}.}
\begin{align}
h_{ij}^{(s)}[t,\vec{x}] &\approx h_{ij}^{(\text{tt})}[t,\vec{x}], \\
h_{ij}^{(\text{tt})}[t,\vec{x}] &\equiv P_{ij ab} h_{ab}^{(\text{de Donder})}[t,\vec{x}] .
\end{align}
Since $P_{ij ab} \delta^{ab} = 0$ -- recall eq. \eqref{GR_ttProjector} -- note that $h_{ij}^{(\text{tt})} = P_{ij ab} \bar{h}_{ab}^{(\text{de Donder})}$ too, because the second term on the right hand side of eq. \eqref{hbToh} would drop out. This in turn implies, from equations \eqref{hbij_EvenDim_NR} and \eqref{hbij_OddDim_NR}:
\begin{align}
h_{ij}^{(s)}[t,\vec{x}] 
\label{hSynch_EvenDim}
&\approx -\frac{8 \pi \GN}{2(2\pi r)^{\frac{d}{2}-1}} \left( \partial_t^{\frac{d}{2}} Q_{ij}^{\text{(tt)}}[t-r] + \mathcal{O}[r^{-1}] \right), \qquad 
\text{(Even $d \geq 4$)}; \\
\label{hSynch_OddDim}
&\approx -\frac{8 \pi \GN}{\sqrt{4\pi} (2\pi r)^{\frac{d}{2}-1}} \int_{0}^{\infty} e^{-\mu\cdot 0^+} \left( \partial_t^{\frac{d+1}{2}} Q_{ij}^{\text{(tt)}}[t-r-\mu]
+ \mathcal{O}[r^{-1}] \right) \frac{\dd\mu}{\sqrt{\mu}}, \qquad
\text{(Odd $d \geq 5$.)} ;
\end{align}
where the transverse-traceless quadrupole has been defined in eq. \eqref{GR_Quadrupole_tt}. 

I now model the finite duration quasi-periodic quadrupole as
\begin{align}
	\label{QuasiPeriodic_Quadrupole_EvenDim}
	\partial_t^{\frac{d}{2}} Q_{ij}[0 \leq t \leq T] 
	&= C^\ell_{ij} \cdot \frac{\sin\left[ \omega_\ell t \right]}{\sqrt{T/2}} ,
	\qquad\qquad \text{(Even $d \geq 4$)} , \\
	\label{QuasiPeriodic_Quadrupole_OddDim}
	\partial_t^{\frac{d+1}{2}} Q_{ij}[0 \leq t \leq T] 
	&= C^\ell_{ij} \cdot \frac{\sin\left[ \omega_\ell t \right]}{\sqrt{T/2}} ,
	\qquad\qquad \text{(Odd $d \geq 3$)} ;
\end{align}
and zero for $t$ outside this interval. The frequency is labeled by an integer $\ell=1,2,3,\dots$ through $\omega_\ell \equiv \pi\ell/T$, and the $\{ C^\ell_{ij} \}$ are arbitrary coefficients. The even $d \geq 4$ synchronous gauge perturbation in eq. \eqref{hSynch_EvenDim} is now
{\allowdisplaybreaks\begin{align}
h_{ij}^{(s)}[t,\vec{x}] 
&\approx -\frac{8 \pi \GN}{2(2\pi r)^{\frac{d}{2}-1}} P_{ij ab} C_{ab}^\ell 
\frac{\sin\left[ \omega_\ell (t-r) \right]}{\sqrt{T/2}}, & 0 \leq t-r \leq T , \\
&= 0 & \text{(Otherwise)} .
\end{align}}
Whereas for the odd $d \geq 5$ case in eq. \eqref{hSynch_OddDim}, borrowing the integration results at the end of \S \eqref{Section_EM},
{\allowdisplaybreaks\begin{align}
h_{ij}^{(s)}[t,\vec{x}] 
&\approx -\frac{8 \pi \GN}{(2\pi r)^{\frac{d}{2}-1}} \frac{P_{ij ab} C_{ab}^\ell}{\sqrt{\omega_\ell T}} \mathcal{E}[\omega_\ell(t-r)] ,
\end{align}}
with
\begin{align}
\mathcal{E}[\omega_\ell T < 0]
	&= 0 \\
\mathcal{E}[0 < \omega_\ell(t-r) < \omega_\ell T]
	&= \sin\left[ \omega_\ell (t-r) \right] \text{FC}\left[ \sqrt{\frac{2\omega_\ell(t-r)}{\pi}} \right] - \cos\left[ \omega_\ell (t-r) \right] \text{FS}\left[ \sqrt{\frac{2\omega_\ell(t-r)}{\pi}} \right] \\
\mathcal{E}[\omega_\ell(t-r) > \omega_\ell T]
	&=
	\sin\left[ \omega_\ell (t-r) \right] \left\{ \text{FC}\left[ \sqrt{\frac{2\omega_\ell(t-r)}{\pi}} \right] - \text{FC}\left[ \sqrt{\frac{2\omega_\ell(t-r-T)}{\pi}} \right] \right\} \\
	&\qquad\qquad\qquad\qquad
	- \cos\left[ \omega_\ell (t-r) \right] \left\{ \text{FS}\left[ \sqrt{\frac{2\omega_\ell(t-r)}{\pi}} \right] - \text{FS}\left[ \sqrt{\frac{2\omega_\ell(t-r-T)}{\pi}} \right] \right\} . \nonumber
\end{align}
The leading asymptotic expansions of $\mathcal{E}$ are
\begin{align}
\mathcal{E}[0 < \omega_\ell(t-r) < \omega_\ell T] 
	&\sim \frac{\sin\left[ \omega_\ell (t-r) - \frac{\pi}{4} \right]}{\sqrt{2}} 
	+ \frac{1}{\sqrt{2 \pi \omega_\ell (t-r)}} \\
\mathcal{E}[\omega_\ell(t-r) > \omega_\ell T]
	&\sim \frac{1}{\sqrt{2 \pi \omega_\ell (t-r)}} 
	- \frac{\cos\left[ \omega_\ell T \right]}{\sqrt{2 \pi \omega_\ell (t-r-T)}}  . 
\end{align}
Since the core of the electromagnetic and gravitational solutions is the massless scalar Green's function, it should not surprise the reader that the fractional distortion of space in eq. \eqref{FractionalDistortion} governed by $h_{ij}^{\text{(tt)}}$ yields a similar time dependence as its real-time electromagnetic power loss counterpart. In even dimensions, the gravitational wave tracks the time dependence of the quadrupole source and is non-zero only when the retarded time $t-r$ lies within the quadrupole's active duration $0 \leq t-r \leq T$. Whereas, in odd dimensions -- at least for sufficiently high frequencies -- the gravitational wave is phase shifted by $-\pi/4$ relative to the quadrupole and is further appended by non-oscillatory signals that fall off as inverse square roots of time. After the retarded time has passed the quadrupole's active duration $t-r>T$, moreover, the now pure tail fractional distortion of space becomes strictly non-oscillatory and decays rapidly to zero. As detected by a hypothetical LIGO or Virgo experiment residing in odd dimensional spacetime, the duration of the quadrupole's active production of gravitational waves is still roughly $T$, despite the presence of the tail effect. Fig. \eqref{LightBulbInOddDim} captures the situation here, just as it did for the electromagnetic case.

\section{Summary and Discussions}
\label{Section_Byebye}

In this work, I have managed to generalize to all relevant dimensions the leading order non-relativistic dipole radiation formula of Maxwell's electromagnetism and the weak field quadrupole gravitational radiation formula of Einstein's General Relativity without a cosmological constant -- $d \geq 3$ for the former and $d \geq 4$ for the latter. The differential energy flux can be found in equations \eqref{Maxwell_DipoleEnergyAngularRadiation_EvenDim} and \eqref{Maxwell_DipoleEnergyAngularRadiation_OddDim} for electromagnetism; and equations \eqref{GR_EnergyFlux_EvenDim_PerOmega} and \eqref{GR_EnergyFlux_OddDim_PerOmega} for weak field $\Lambda=0$ General Relativity. Their corresponding integrated power loss can be found in \S \eqref{Section_Introduction}. For angular momentum radiated to infinity, the electromagnetic results can be found in equations \eqref{Maxwell_AngularMomentumFlux_EvenDim_PerOmega}, \eqref{Maxwell_AngularMomentumFlux_EvenDim_Total}, \eqref{Maxwell_AngularMomentumFlux_OddDim_PerOmega}, and \eqref{Maxwell_AngularMomentumFlux_OddDim_Total}. Whereas, the gravitational wave angular momentum loss in equations \eqref{GR_AngularMomentumFlux_EvenDim} and \eqref{GR_AngularMomentumFlux_OddDim}. The main novelty uncovered here is the dependence of the radiation on the entire past histories of the dipoles and quadrupoles in odd dimensional Minkowski due to the tail effect -- to my knowledge, these real-time expressions have not appeared before in the literature. This stark distinction between the causal structure of massless signals in odd versus even dimensions is lost in their frequency space expressions, where they take the same form for all dimensions.

Despite this tail-induced history dependence in odd dimensions, however, I have also shown that -- apart from a phase shift and additional non-oscillatory terms that decays as inverse square roots of time -- a (quasi-)periodic electromagnetic or gravitational source of duration $T$ remains roughly the same duration $T$ when seen or heard by a distant observer. It may be interesting to investigate more complex radiating systems, where the tail effect may exhibit other features that would further distinguish it from tail-free propagation in even dimensions.

While the conceptual understanding of the tail effect as inside-the-null-cone propagation allowed us to anticipate that the far zone radiation has to depend on the entire past history of its sources, the key technical step involved the integral representation of the $\Gamma$ function in eq. \eqref{GammaFunction}, an insight I borrowed from quantum field theory. This converted the inverse fractional powers of frequencies occurring in the odd dimensional Green's functions -- absent in the even case -- into a one dimensional integral over $\mu$ involving the outgoing wave $\exp[i\omega(t-r+\widehat{r}\cdot\vec{x}'-\mu)]$; see eq. \eqref{G_OddDim_FarZone_v4}. In other words, the integral representation of $\Gamma$ is in fact the integral over the past history. For future work, it may be illuminating to re-derive the odd dimensional power loss results from the source point of view; i.e., by energy conservation, the work done on the source by the history-dependent electromagnetic and gravitational forces must in fact be equal to the power loss to infinity. Instead of the far zone expansions I carried out in this paper, such an analysis would require a near zone one instead. The techniques deployed could in turn be potentially relevant for other self-force calculations.

To be sure, our gravitational results were derived with the assumption that the energy-momentum-shear-stress of matter is conserved. This excludes self-gravitating systems such as the compact binaries whose gravitational waves LIGO and Virgo have been hearing from; for such a scenario it is the stress energy tensor of both matter and gravitation that is conserved -- even though the final answers are likely to be the same, I hope in the near future to re-derive the quadrupole formulas for the self-gravitation case for all $d \geq 4$. Curiously, even gravitation textbooks and review articles do not always treat this issue with care. For instance, some assume the stress energy tensor of matter is conserved, but go on to claim that the formulas must be valid for the compact binary system. I should mention, the quantum field theory techniques used in \cite{Cardoso:2008gn} does take into account the gravitational dynamics of the compact binary system, and their results are therefore valid for self-gravitating systems even though they were not able to capture the real-time power loss.

I end with a remark on the far zone radiation limit and its relation to large number of dimensions $d$. For the radiation signal -- the $1/r^{(d/2)-1}$ piece of the field that contributes to energy loss to infinity -- to be cleanly separated from all the subleading orders in the $1/r$ expansion, it must be dominant in magnitude over the rest. As an estimate, let us examine the first subleading terms (which scale as $1/r^{d/2}$) in equations \eqref{G_EvenDim_FarZone_v3} and \eqref{G_OddDim_FarZone_v3}, neglecting the $\omega \widehat{r}\cdot\vec{x}' = (\omega r)(\widehat{r} \cdot\vec{x}'/r) \ll \omega r$ terms. At large dimensions, this radiation condition translates to $\omega r \gg n^2/2 \sim d^2/8$. For any detector placed at finite distances from the source, therefore, we infer that the notion of radiation may become increasingly ambiguous with extremely large number of spacetime dimensions.

\section{Acknowledgments}

I thank Wei-Hao Chen, Meng-Han Kuo, and Cheng-Han Wu for initial collaborations on this work. I also wish to acknowledge useful discussions with Yen-Wei Liu, as well as email exchanges with D.  Gal’tsov and M. Khlopunov. Much of the computations in this paper was carried out with the aid of {\sf Mathematica} \cite{MMA} and the tensor package {\sf xAct} \cite{xAct}. The references \cite{DLMF} and \cite{GnS} were consulted repeatedly for the relevant properties of special functions. I was supported by the Ministry of Science and Technology of the R.O.C. under the grant 106-2112-M-008-024-MY3.

\appendix

\section{Solid Angle Integrals}
\label{Section_Calculus}

We wish to evaluate the following tensor integral over the $(d-2)-$dimensional round sphere $\mathbb{S}^{d-2}$:
\begin{align}
\label{TensorIntegral}
I_{\mathbb{S}^{d-2}}^{i_1 \dots i_N}
\equiv \int_{\mathbb{S}^{d-2}} \dd^{d-2}\Omega \widehat{r}^{i_1} \widehat{r}^{i_2} \dots \widehat{r}^{i_N} .
\end{align}
We may first observe that under parity, every unit radial vector reverses sign, $\widehat{r}^{i_1} \dots \widehat{r}^{i_N} \to (-)^N \widehat{r}^{i_1} \dots \widehat{r}^{i_N}$, while the solid angle measure is invariant. This implies the tensor integral is zero for all odd $N \geq 1$. Next, notice
\begin{align}
\label{TensorIntegral_GeneratingFunction}
I_{\mathbb{S}^{d-2}}^{i_1 \dots i_N}
= \left. \frac{1}{i^N} \frac{\partial^N}{\partial k^{i_1} \dots \partial k^{i_N}} \right\vert_{\vec{k}=\vec{0}}
\int_{\mathbb{S}^{d-2}} \dd^{d-2}\Omega e^{i\vec{k}\cdot\widehat{r}} .
\end{align}
The scalar integral $I_{\mathbb{S}^{d-2}} \equiv \int \dd^{d-2}\Omega e^{i\vec{k}\cdot\widehat{r}}$ is thus a generating function for the tensor integral $I_{\mathbb{S}^{d-2}}^{i_1 \dots i_N}$, and may be tackled by first utilizing the spherical symmetry of the problem to set $\vec{k} = k \widehat{e}_{d-1}$, where $\widehat{e}_{d-1}$ is the unit vector along the $(d-1)$th spatial axis, so that $\vec{k} \cdot \widehat{r} \equiv k \cos\theta^{d-2}$, with $\theta^{d-2}$ denoting the $(d-2)$th angle parametrizing the round $(d-2)-$sphere. Moreover, the solid angle measure $\dd\Omega$ in $d-1$ and $d-2$ spatial dimensions are related via
\begin{align}
\label{SolidAngle_Recursion}
\dd^{d-2}\Omega
= \dd (\cos\theta^{d-2}) (1-\cos^2\theta^{d-2})^{\frac{d-4}{2}} \cdot \dd^{d-3}\Omega .
\end{align}
Using the integral representation of the Beta function $B[\alpha,\beta] = \Gamma[\alpha]\Gamma[\beta]/\Gamma[\alpha+\beta]$, one may then show by repeated iteration of eq. \eqref{SolidAngle_Recursion} that the solid angle in $d-1$ spatial dimensions is
\begin{align}
	\label{SolidAngle}
	\int_{\mathbb{S}^{d-2}} \dd^{d-2}\Omega
	&= \frac{2\pi^{\frac{d-1}{2}}}{\Gamma[\frac{d-1}{2}]} .
\end{align}
Altogether, $\int \dd^{d-2}\Omega e^{i\vec{k}\cdot\widehat{r}}$ is thus the solid angle subtended by a $d-3$ dimensional round sphere, multiplied by the integral involving the final angular variable $\theta^{d-2}$:
\begin{align}
\int_{\mathbb{S}^{d-2}} \dd^{d-2}\Omega e^{i\vec{k}\cdot\widehat{r}}
&= \frac{2 \pi^{\frac{d-2}{2}}}{\Gamma[\frac{d-2}{2}]} \int_{-1}^{+1} \dd c(1-c^2)^{\frac{d-4}{2}} \exp\left[ i k c \right] \\
\label{TensorIntegral_GeneratingFunction_Step1}
&= 2 \pi^{\frac{d-1}{2}} \sum_{s=0}^{\infty} \frac{i^{2s} (\delta_{ab} k^a k^b)^s}{4^s \cdot s! \Gamma[\frac{d-3}{2}+s+1]} .
\end{align}
In the second line, we have further recognized the integral representation of the Bessel function
\begin{align}
	\int_{-1}^{+1} \dd c (1-c^2)^{\nu - \frac{1}{2}} \exp[icz]
	= \frac{\sqrt{\pi} \Gamma[\nu + \frac{1}{2}]}{(z/2)^\nu} J_\nu[z] , 
	\qquad
	\text{Re}[z] > -\frac{1}{2} ;
\end{align}
and hence proceeded to use its power series representation,
\begin{align}
	J_\nu[z]
	&= \left( \frac{z}{2} \right)^\nu
	\sum_{\ell=0}^{\infty} \frac{(iz)^{2\ell}}{2^{2\ell} \ell! \Gamma[\nu+\ell+1]} .
\end{align}
Differentiating the $i^{2s} \vec{k}^{2s}$ factor in eq. \eqref{TensorIntegral_GeneratingFunction_Step1} $N=2\ell$ times -- cf. eq. \eqref{TensorIntegral_GeneratingFunction} -- and setting $\vec{k}=\vec{0}$ would yield a non-zero result only from the $s=\ell$ term of the summation. Furthermore, we may phrase the ensuing expression as a `sum-over-contractions', to borrow quantum field theory lingo. Define a contraction between a pair of $k$'s by replacing them with the corresponding Kronecker delta; for instance, contraction of $k^i k^j$ yields $\delta^{ij}$. Then, I claim that
\begin{align}
\label{DifferentiationAsContractions}
\frac{\partial^{2\ell}}{\partial k^{i_1} \dots \partial k^{i_{2\ell}}} k^{2\ell}
	= 2^\ell \cdot \ell! \sum \left(\text{Full contractions of }k^{i_1} \dots k^{i_{2\ell}}\right) .
\end{align}
The first few cases are
{\allowdisplaybreaks\begin{align}
\frac{\partial^{2}}{\partial k^{i_1} \partial k^{i_{2}}} k^{2}
&= 2 \delta^{i_1 i_2}, \\
\frac{\partial^{4}}{\partial k^{i_1} \dots \partial k^{i_{4}}} k^{4}
&= 8 \left( \delta^{i_1 i_2} \delta^{i_3 i_4} + \delta^{i_1 i_3} \delta^{i_2 i_4} + \delta^{i_1 i_4} \delta^{i_2 i_3} \right), \\
\frac{\partial^{6}}{\partial k^{i_1} \dots \partial k^{i_{6}}} k^{6}
&= 48 ( \delta^{i_1 i_2} \delta^{i_3 i_4} \delta^{i_5 i_6} + \delta^{i_1 i_3} \delta^{i_2 i_4} \delta^{i_5 i_6} + \delta^{i_1 i_4} \delta^{i_2 i_3} \delta^{i_5 i_6} \nonumber\\
&\qquad\qquad
\delta^{i_1 i_2} \delta^{i_3 i_5} \delta^{i_4 i_6} + \delta^{i_1 i_3} \delta^{i_2 i_5} \delta^{i_4 i_6} + \delta^{i_1 i_5} \delta^{i_2 i_3} \delta^{i_4 i_6} \nonumber\\
&\qquad\qquad
\delta^{i_1 i_2} \delta^{i_5 i_4} \delta^{i_3 i_6} + \delta^{i_1 i_5} \delta^{i_2 i_4} \delta^{i_3 i_6} + \delta^{i_1 i_4} \delta^{i_2 i_5} \delta^{i_3 i_6} \nonumber\\
&\qquad\qquad
\delta^{i_1 i_5} \delta^{i_3 i_4} \delta^{i_2 i_6} + \delta^{i_1 i_3} \delta^{i_5 i_4} \delta^{i_2 i_6} + \delta^{i_1 i_4} \delta^{i_5 i_3} \delta^{i_2 i_6} \nonumber\\
&\qquad\qquad
\delta^{i_5 i_2} \delta^{i_3 i_4} \delta^{i_1 i_6} + \delta^{i_5 i_3} \delta^{i_2 i_4} \delta^{i_1 i_6} + \delta^{i_5 i_4} \delta^{i_2 i_3} \delta^{i_1 i_6} ) .
\end{align}}
To prove this assertion, let us write $k^{2\ell}$ in terms of contractions with the appropriate Kronecker deltas.
\begin{align}
\label{DifferentiationAsContractions_Step1}
\frac{\partial^{2\ell}}{\partial k^{i_1} \dots \partial k^{i_{2\ell}}} k^{2\ell}
&= \frac{\partial^{2\ell}}{\partial k^{i_1} \dots \partial k^{i_{2\ell}}} \left( k^{j_1} \dots k^{j_{2\ell}} \delta_{j_1 j_2} \delta_{j_3 j_4} \dots \delta_{j_{2\ell-1} j_{2\ell}} \right) .
\end{align}
Each derivative acting on one of the $k$'s would simply transfer the index on the derivative onto the Kronecker delta that the particular $k$ is contracted with; for example, $\partial_{k^a} k^b \delta_{bc} = \delta_{ac}$, the $a$ is transferred onto the Kronecker delta. But by the product rule, each derivative would act on each $k$ once, and therefore the result of eq. \eqref{DifferentiationAsContractions_Step1} must simply be the product of $\ell$ Kronecker deltas, summed over all permutations of their indices evaluated on the set $\{i_1, \dots, i_{2\ell}\}$. But since the Kronecker deltas are symmetric rank 2 tensors, and there are $\ell$ of them in each product, there must be $2^\ell \cdot \ell!$ identical terms. This proves eq. \eqref{DifferentiationAsContractions}.

{\bf Results} \qquad Finally, applying eq. \eqref{DifferentiationAsContractions} to equations \eqref{TensorIntegral_GeneratingFunction} and \eqref{TensorIntegral_GeneratingFunction_Step1} now returns
\begin{align}
\label{TensorIntegral_EvenPointFunction}
\int_{\mathbb{S}^{d-2}} \dd^{d-2}\Omega \widehat{r}^{i_1} \dots \widehat{r}^{i_{2\ell}}
&= \frac{\pi^{\frac{d-1}{2}}}{2^{\ell-1} \Gamma[\frac{d-3}{2}+\ell+1]}
\sum \left(\text{Full contractions of }k^{i_1} \dots k^{i_{2\ell}}\right) .
\end{align}
We also collect here the earlier parity-based argument, that these tensor integrals are zero whenever there are odd powers of $\widehat{r}$.
\begin{align}
\label{TensorIntegral_OddPointFunction}
\int_{\mathbb{S}^{d-2}} \dd^{d-2}\Omega \widehat{r}^{i_1} \dots \widehat{r}^{i_{2\ell}} \widehat{r}^{i_{2\ell+1}}
&= 0, \qquad\qquad \ell = 0,1,2,\dots .
\end{align}
The two cases used in the body of this paper are
\begin{align}
\label{TensorIntegral_2PointFunction}
\int_{\mathbb{S}^{d-2}} \dd^{d-2}\Omega \widehat{r}^{i_1} \widehat{r}^{i_2}
	&= \frac{\pi^{\frac{d-1}{2}}}{\Gamma[\frac{d+1}{2}]} \delta^{i_1 i_2} , \\
\label{TensorIntegral_4PointFunction}
\int_{\mathbb{S}^{d-2}} \dd^{d-2}\Omega \widehat{r}^{i_1} \widehat{r}^{i_2} \widehat{r}^{i_3} \widehat{r}^{i_4}
	&= \frac{\pi^{\frac{d-1}{2}}}{2 \Gamma[\frac{d+3}{2}]}
	\left( \delta^{i_1 i_2} \delta^{i_3 i_4} + \delta^{i_1 i_3} \delta^{i_2 i_4} + \delta^{i_1 i_4} \delta^{i_2 i_3} \right) .
\end{align}

\end{document}